\documentclass[traditabstract]{aa}

\usepackage[T1]{fontenc}
\usepackage{ae,aecompl}
\usepackage{graphicx}
\usepackage{txfonts}
\usepackage{amsmath}
\usepackage{amssymb}
\usepackage{sidecap}
\usepackage{xcolor}
\usepackage{subfig}
\usepackage{appendix}
\usepackage{rotating}
\bibpunct{(}{)}{;}{a}{}{,} 

\usepackage{hyperref}
\usepackage{upgreek}

\hypersetup{
  colorlinks   = true, 
  urlcolor     = blue, 
  linkcolor    = red, 
  citecolor    = blue 
}

\newcommand{\asec}{^{\prime\prime}}

\begin{document}

\title{Constraining the AGN duty cycle in the cool-core cluster\\ MS 0735.6+7421 with LOFAR data}
\author{Nadia Biava 
            \inst{1,2}
            \and
            Marisa Brienza 
            \inst{1,3} 
            \and
            Annalisa Bonafede
            \inst{1,3}
            \and
            Myriam Gitti
            \inst{1,3}
            \and
            Etienne Bonnassieux
            \inst{1,3}
            \and
        Jeremy Harwood
            \inst{4}
            \and
            Alastair C. Edge
            \inst{5}
            \and
            Christopher J. Riseley
            \inst{1,3}
            \and
            Adrian Vantyghem
            \inst{6}
}

   \institute{Dipartimento di Fisica e Astronomia, Università di Bologna, via P. Gobetti 93/2, I-40129, Bologna, Italy
         \and 
             INAF - Osservatorio di Astrofisica e Scienza dello Spazio di Bologna, via Gobetti 93/3, I-40129 Bologna, Italy
         \and 
             INAF - Istituto di Radioastronomia, Bologna Via Gobetti 101, I-40129 Bologna, Italy        
        \and
            Centre for Astrophysics Research, University of Hertfordshire, College Lane, Hatfield AL10 9AB, UK 
        \and
            Centre for Extragalactic Astronomy, Durham University, DURHAM, DH1 3LE, UK 
        \and
            University of Manitoba, Department of Physics and Astronomy, Winnipeg, MB R3T 2N2, Canada
             }

   \date{Received ; accepted }

\abstract 
{\emph{Context.} MS 0735.6+7421 is a galaxy cluster that hosts a central radio galaxy with a very steep spectrum. The spectrum is produced by one of the most powerful known jetted active galactic nuclei (AGN). The radio plasma, ejected at nearly light speed from the central AGN, has displaced the intra-cluster medium, leaving two pairs of cavities observable in the X-ray. The cavities are associated with two different outbursts and have distributed energy to the surrounding medium.
While the age of the cavities has previously been estimated from the X-rays, no confirmation from radio data is available. 
Furthermore, the radio spectrum has only been derived from integrated flux density measurements so far, and the spatial distribution that would help us to understand the nature of this source is still lacking.
\\\emph{Aims.} We perform for the first time a detailed, high-resolution spectral study of the source at radio frequencies and investigate its duty cycle. We compare this with previous X-ray estimates.
\\\emph{Methods.} We used new observations at 144 MHz produced with the LOw Frequency ARray (LOFAR) together with archival data at higher frequencies (235, 325, 610, 1400, and 8500 MHz), to investigate the spectral properties of the source. We also used radiative models to constrain the age of the source.
\\\emph{Results.}
At the LOFAR frequency, the source presents two large outer radio lobes that are wider than at higher frequencies, and a smaller intermediate lobe that is located south-west of the core. A new inspection of X-ray data allowed us to identify an intermediate cavity that is associated with this lobe. It indicates a further phase of jet activity. The radio lobes have a steep spectrum even at LOFAR frequencies, reaching $\alpha_{144}^{610}=2.9$ in the outer lobes and $\alpha_{144}^{610}=2.1$ in the intermediate lobe. Fitting the lobe spectra using a single injection model of particle ageing, we derived a total age of the source between 170 and 106 Myr. This age agrees with the buoyancy and sound-crossing timescales derived from X-ray data. The resolution of the spectral age map we performed allows us to reconstruct the duty cycle of the source. In three phases of jet activity, the AGN was active for most of the time with only brief quiescent phases that ensured the repeated heating of the central gas. Finally, we estimated the minimum energy inside the outer lobes. We find that a source of additional pressure support must be present to sustain the bubbles against the pressure of the external medium.}

\titlerunning{}
\maketitle

\section{Introduction}

Galaxy clusters are the largest gravitationally bound objects in the Universe, with masses up to $\sim 10^{15}\ \rm{M_{\odot}}$. They host hundreds to thousands of galaxies.
The space in between the galaxies is filled with a hot ($T\sim 10^7 - 10^8 \, \rm{K}$) and rarefied gas (electron density $n_e \sim 10^{-3} \, \rm{cm^{-3}}$) that is known as the intra-cluster medium (ICM), which emits in the X-ray band through bremsstrahlung radiation \citep[e.g.][]{Mitchell1976, Serlemitsos1977, Forman1982}. Radio observations reveal relativistic particles and magnetic field in the ICM \citep[see e.g. the review by][]{vanWeeren2019}.

Galaxy clusters often host active galactic nuclei (AGN) with jets of relativistic plasma, which emit synchrotron radiation and are mostly visible at radio frequencies \citep[e.g.][]{deYoung1984, deYoung2002, Tadhunter2016}. 
The sizes of these sources range from a few kiloparsec to a few magaparsec and extend well beyond the host galaxy. 

The hot X-ray emitting gas in the central region of many cool-core clusters has a radiative cooling time that is much shorter than the Hubble time. Therefore a cooling flow is expected to develop \citep{Fabian1994}.
In the standard model of cooling flows, the gas cools from the cluster temperature down to temperatures significantly below 1 keV, emitting different atomic lines. However, high-resolution spectroscopic observations with the \emph{Chandra} and XMM-\emph{Newton} X-ray telescopes have not found evidence of the predicted cooling to low temperatures, implying that some source of heating must balance the radiative losses \citep{Peterson2003, PetersonFabian2006}.

The most plausible heating mechanism is mechanical feedback from the AGN hosted by the cluster, in particular, by the brightest cluster galaxy \citep[BCG; see e.g. the reviews by][]{McNamaraNulsen2007, McNamaraNulsens2012, HardcastleCroston2020}. Evidence of this hypothesis is found in the X-ray cavities observed in a number clusters, on scales approximately coincident with the lobes of the central radio galaxy \citep[e.g.][]{McNamara2000, Fabian2000}.
In this scenario, the radio plasma emitted by the central AGN displaces the X-ray emitting gas, creating a low-density bubble that rises buoyantly and expands, distributing energy to the surrounding medium. 
To suppress cooling over the age of clusters, repeated events of AGN jet activity are required to occur on timescales shorter than the cooling time.

In some objects, two systems of cavities have been observed \citep[e.g.][]{Nulsen2005,Wise2007,Vantyghem2014} or the AGN shows multiple radio lobes that are related to repeated outbursts \citep[][]{Morganti2017}, sometimes pointing in different directions \citep[e.g. RBS797;][]{Doria2012, Gitti2006}.
For restarted sources, the spectral properties of the radio plasma can be used to derive their duty cycle, that is, the fraction of time during which the AGN is active \citep[e.g.][]{Harwood2013,Harwood2015,Brienza2020}.

However, performing resolved spectral studies to derive accurate estimates of the age of the AGN plasma is not trivial as it requires high-resolution and sensitive radio observations over a wide range of frequencies. The advent of instruments such as the LOw Frequency ARray \citep[LOFAR;][]{vanHaarlem2013} allows us for the first time to perform these studies at low frequencies.

\subsection{MS 0735.6+7421}
One of the most famous examples of AGN mechanical feedback is represented by the radio source located at the centre of the cool-core galaxy cluster MS 0735.6+7421 (hereafter MS0735). The source (RA = 07:41:40.3, DEC = +74:14:58.0) has a redshift of $z = 0.216$, corresponding to a scale of 1$\asec$ = 3.53 kpc (assuming a $\Lambda$CDM cosmological model with $\Omega_{\Lambda} = 0.7$, $\Omega_m = 0.3,$ and $H_0 = 70\ {\rm km\ s^{-1}\ Mpc^{-1}}$). 

This cluster was first identified as a cooling flow candidate by \cite{Donahue1992} and was then confirmed by \cite{DonahueStocke1995}. Based on data obtained with the Very Large Array (VLA), the central radio source (4C +74.13) was identified as a radio galaxy with a clearly defined core and outer lobes \citep{Cohen2005}. Cohen et al. (2005) also found a significant extension of the central emission towards the south-west, which may be the result of the jet interacting with the ICM, or an inner lobe generated by a more recent radio outburst. Using VLA data at 325 MHz and 1425 MHz, with a resolution of $21\asec$, the authors found steep spectra in all the main features of the source: $\alpha_{325}^{1425} = 1.54 \pm 0.04$ for the central region, and $\alpha_{325}^{1425} = 3.17 \pm 0.05$ and $\alpha_{325}^{1425} = 3.13 \pm 0.05$ for the northern and southern outer lobes, respectively (where $S_{\rm \nu}\propto \nu^{-\alpha}$). 

\emph{Chandra} X-ray observations revealed two giant cavities, each roughly 200 kpc in diameter, and a shock front surrounding them \citep{McNamara2005}. The radio lobes partially fill the cavities, suggesting that the gas was displaced and compressed by the advancing radio jets. The total energy involved in the process is $\sim 6 \times 10^{61}$ erg, making this one of the most powerful known outbursts, together with the Ophiuchus galaxy cluster \citep{Giacintucci2020}. The total energy is enough to quench the cooling flow and to heat the gas within 1 Mpc by $\sim 0.25$ keV per particle. \emph{XMM} data confirmed these results and showed that most cooling-flow clusters are likely experiencing such powerful outbursts during episodes lasting for a small fraction ($\sim 10\%$) of their age \citep{Gitti2007}.

Using deep \emph{Chandra} observations, \cite{Vantyghem2014} detected a second pair of cavities located along the radio jets in the inner $20$ kpc, which corroborated the hypothesis of a renewed period of AGN activity. The authors estimated the cavity ages using the buoyancy, sound crossing, and refill timescales. The mean ages of the northern and southern outer cavities are 150 Myr and 170 Myr, respectively, while the age of the surrounding shock front (110 Myr), which should be comparable to the true cavity age, is shorter than the mean rise time. This means that the cavity ages are slightly overestimated while the power to inflate the bubbles (for the northern cavity, $9\times10^{45}\ \rm{erg\ s^{-1}}$; for the southern cavity, $8\times10^{45}\ \rm{erg\ s^{-1}}$) is underestimated.
For the inner cavities, Vantyghem and collaborators found a mean age of 48 Myr (northern cavity) and 42 Myr (southern cavity). The power of these cavities (northern $2.8\times10^{44}\ \rm{erg\ s^{-1}}$ and southern $2.4\times10^{44}\ \rm{erg\ s^{-1}}$) is 30 times smaller than the outer cavity power, implying that the jet power has varied significantly over time \citep[][]{Vantyghem2014}. The outburst interval, estimated as the difference between the ages of the outer and inner cavities, is 110 Myr. It is shorter than the central cooling time, which prevents most of the surrounding gas from cooling. Repeated heating of the central gas supplies enough energy to prevent cooling, and may explain the lack of star formation in the system \citep{McNamara2009}.

\begin{figure}
    \resizebox{\hsize}{!}{\includegraphics{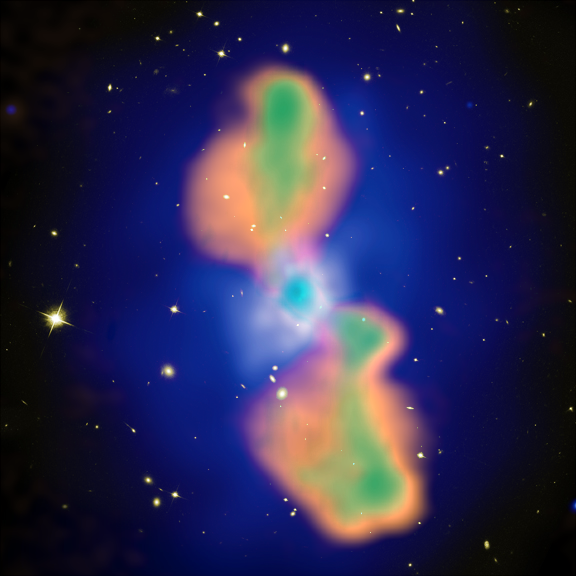}}
    \caption{Composite image of MS0735, obtained by combining the \emph{Chandra} X-ray image (blue), the I-band image taken by the Hubble Space Telescope (white), the radio wavelengths observed with LOFAR at 144 MHz (orange) and those observed with the VLA at 325 MHz (green). The LOFAR lobes are wider than at 325 MHz, and they perfectly fill the cavities.}
    \label{fig:Composite}
\end{figure}

However, the ages derived from the cavity rise time are approximate and are an indirect probe of the AGN duty cycle. A first estimate of the radiative age of the lobes from radio data was performed by \cite{Birzan2008}. Based on integrated spectral analysis of VLA data at 235 MHz and 1400 MHz, they estimated a synchrotron age of 97 Myr, which reduces to 50 Myr if adiabatic losses are considered.

Integrated models of spectral ageing have long been used to determine the age of radio galaxies. However, they provide a poor, frequency-dependent description of the spectrum of a source compared to recent well-resolved studies that use modern analysis techniques on small spatial scales \citep{Harwood2017a}.

In this paper, we present new radio observations of MS0735 that were performed with LOFAR at 144 MHz. A composite image of the source, created with our new LOFAR data together with archival VLA 325 MHz data, is shown in Fig. \ref{fig:Composite}. The high resolution and sensitivity provided by LOFAR at 144 MHz enable us to characterize the morphology, spectral properties, and radiative age of this source in detail for the first time.
Our aim is to perform a resolved spectral study of MS0735 for which we combine LOFAR data with archival observations at higher frequencies to investigate the duty cycle of the central AGN and its energetics.

We note that LOFAR snapshot observations of MS0735 have previously been shown by \cite{Kokotanekov2017}. This observation was performed during the first LOFAR all-sky survey, the Multifrequency Snapshot Sky Survey \citep[MSSS; see][]{Heald15}. The MSSS image of MS0735 has a resolution of $27.8\asec$ and an rms noise of 30 mJy beam$^{-1}$ at 140 MHz, neither of which is sufficient for a resolved spectral study.

The paper is organised as follows: in Sect. \ref{sec:data} we describe the data used in our study and the techniques we used to reduce them. We present our results in Sect. \ref{sec:results} and discuss them in Sect. \ref{sec:discussion}, while our conclusions are reported in Sect. \ref{sec:conclusions}. 
Throughout the paper the spectral index, $\alpha$, is defined using the convention $S\propto \nu^{-\alpha}$.

\begin{table*}[]
    \centering
    \caption{Summary of the observational details.}
    \renewcommand\arraystretch{1.2}
    \begin{tabular}{cccccc} \hline
         Telescope &Frequency &Configuration &Target TOS &Calibrators &Observation date  \\\hline
         LOFAR &144 MHz &HBA-dual inner &8h &3C196 &23 May 2018\\
         GMRT &235 MHz &- &6h &3C147, 3C286, 0834+555 &4 December 2009\\
         VLA &325 MHz &A &2.4h &3C286, 0749+743 &24 December 2004 \\
         VLA &325 MHz &B &3.6h &3C147, 0749+743 &4 December 2003 \\
         VLA &325 MHz &C &4h &3C286, 3C184 &24 December 2006\\
         GMRT & 610 MHz &- &6h &3C147, 3C286, 0834+555 &4 December 2009\\
         VLA* &1420 MHz &A &2h &3C286, 0749+743 &24 October 2004 \\
         VLA &1420 MHz &B &5h &3C286, 0954+745 &19 August 2006\\
         VLA &1420 MHz &C &3.4h &3C147, 0841+708 &15 April 2004\\
         VLA* &8460 MHz &A &1.7h &3C286, 0749+743 &17 October 2004\\
         VLA* &8460 MHz &D &4.5h &3C147, 0721+713 &28 November 2005\\
         \hline
    \end{tabular}
    \tablefoot{*Observations considered only for the analysis of the central emission.}
    \label{tab:obs}
\end{table*}

\begin{table*}[]
    \centering
    \renewcommand\arraystretch{1.2} 
     \caption{Image properties.}
    \begin{tabular}{cccccc} \hline
         Telescope &$\nu$ &uv-range &Beam size &Rms &Fig.\\
         &[MHz] &[$\lambda$]&&[$\rm{mJy\ beam^{-1}}$] &\\\hline
         LOFAR &144 &$48\sim480000$ &7.0"$\times$4.5" &0.2 &\ref{fig:lofar}\\
         GMRT &235 &$30\sim 19173$ &20"$\times$12" &2.7 &\ref{fig:all_images}\\
         VLA &325 &$37\sim 40186$ &7"$\times$5" &0.53 &\ref{fig:all_images}\\
         GMRT &610 &$80\sim 49375$ &7"$\times$4" &$6.0\times10^{-2}$ &\ref{fig:all_images}\\
         VLA &1420 &$157\sim 54376$ &10"$\times$10" &$4.2\times10^{-2}$ &\ref{fig:all_images}\\
         VLA &8460 &$807\sim 29186$&9"$\times$7" &$1.2\times10^{-2}$ &\ref{fig:all_images}\\\hline
    \end{tabular}
    \label{tab:image_properties}
\end{table*}

\section{Data}\label{sec:data}
\begin{figure*}[h]
   \sidecaption
    \includegraphics[width=12cm]{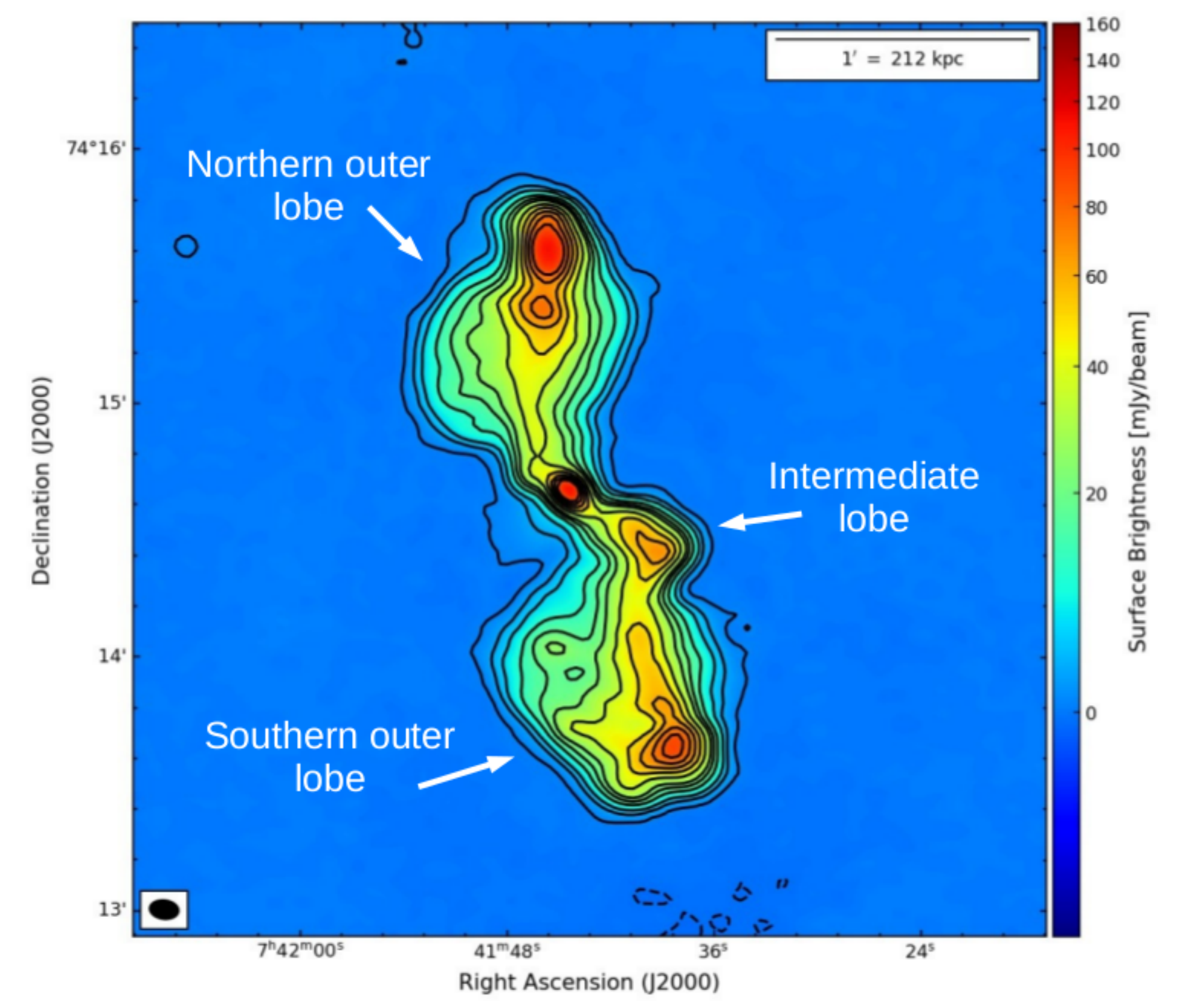}
    \caption{LOFAR 144 MHz radio map of MS0735 at resolution of $7.0\asec\times4.5\asec$. Contour levels: [$-1$,1,5,10,20,30,40,60,80,100,120,140] $\times\ 3\sigma$ (where $\sigma = 0.2\ \rm{mJy\ beam^{-1}}$). The beam is shown in the bottom left corner.}
    \label{fig:lofar}
\end{figure*}

\begin{figure*}[]
\centering
    \includegraphics[width=17cm]{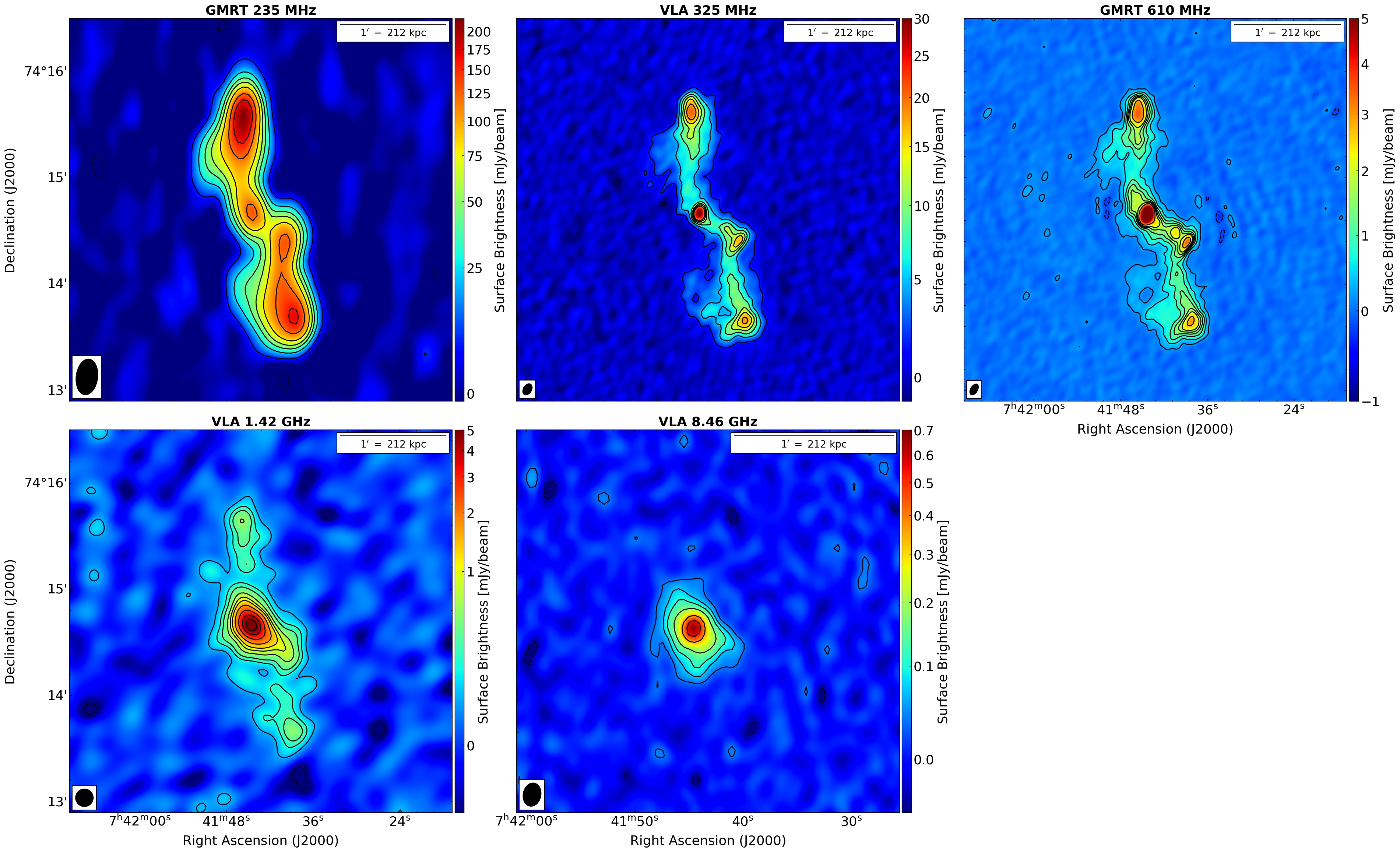}
    \caption{Radio maps of MS0735 at different frequencies. \emph{Top left}: GMRT 235 MHz at $20\asec\times12\asec$ resolution. Levels: [$-1$, 1, 3, 5, 7, 10, 15, 20, 30] $\times\ 3\sigma$ (where $\sigma = 2.7\ \rm{mJy\ beam^{-1}}$). \emph{Top middle}: VLA 325 MHz image at $7\asec\times5\asec$ resolution. Levels: [$-1$, 1, 3, 5, 7, 9, 11, 15, 20] $\times\ 3\sigma$ ($\sigma = 0.53\ \rm{mJy\ beam^{-1}}$). \emph{Top right}: GMRT 610 MHz at $7\asec\times4\asec$ resolution. Levels: [$-1$, 1, 3, 5, 7, 9, 11, 15, 18, 30, 50] $\times\ 3\sigma$ ($\sigma = 6.0\times10^{-2}\ \rm{mJy\ beam^{-1}}$). \emph{Bottom left}: VLA 1420 MHz at $10\asec\times10\asec$ resolution. Levels: [$-1$, 1, 2, 3, 5, 7, 9, 13, 20, 30, 40] $\times\ 3\sigma$ ($\sigma = 4.2\times10^{-2}\ \rm{mJy\ beam^{-1}}$). \emph{Bottom right}: VLA 8460 MHz at $9\asec\times7\asec$ resolution. Levels: [$-1$, 1, 2, 3, 5, 10, 15] $\times\ 3\sigma$ ($\sigma = 1.18\times10^{-2}\ \rm{mJy\ beam^{-1}}$). The beam is shown in the bottom left corner of each image. }
    \label{fig:all_images}
\end{figure*}

To investigate the spectral properties and estimate the radiative age of the radio lobes of MS0735, we used new 144 MHz LOFAR data together with archival data from the VLA and the Giant Metrewave Radio Telescope \citep[GMRT;][]{Swarup1990}, covering the frequency range 144$-$8460 MHz.
We reprocessed archival observations performed with the VLA at 325, 1420, and 8460 MHz and the GMRT at 235 and 610 MHz.
In the following sections, we describe these observations and the data reduction procedures. A summary of the radio observational details is reported in Table \ref{tab:obs}. We present the final images throughout our frequency range in Figs. \ref{fig:lofar} and \ref{fig:all_images}. Their properties are listed in Table \ref{tab:image_properties}.

Finally, we compare the radio morphology of the source with X-ray features, shown in Fig. \ref{fig:X}, and the radiative age of the radio lobes with X-ray cavity timescales, using \emph{Chandra} X-ray observations of this source presented in \cite{Vantyghem2014}. We describe these observations only briefly.

\subsection{LOFAR observation and data reduction}
The source was observed on 23 May 2018 using the LOFAR High Band Antennas (HBA) in the \verb|HBA_DUAL_INNER| mode. The target was observed for a total integration time of 8 h, using all the LOFAR stations: core stations, remote stations, and international stations. International stations were not used in this work and were flagged before calibration. The observations were performed using a dual-beam mode setup, co-observing with the LOFAR Two-Metre Sky Survey \citep[LoTSS;][]{Shimwell2017}. We observed in the frequency range 110$-$190 MHz, divided into 244 sub-bands of 64 channels each, with a sampling time of 1 s. Other details, such as the list of calibrators, are reported in Table \ref{tab:obs}.

This observation was carried out in co-observing mode with the LOFAR Surveys Key Science project. Data were calibrated using the direction-independent (DI) pipeline described in \cite{Shimwell2017} and the direction-dependent (DD) pipeline that uses \verb|DDF| and \verb|killMS| \citep{Tasse2018,Tasse2014a,Tasse2014b,SmirnovTass2015}, following the same approach as was used for the LoTSS 2nd data release \citep[][]{Shimwell2019}. 

Sources outside $30\arcmin$ from the target were subtracted using the DD gains found by \verb|killMS|, and the extracted field was then self-calibrated further and imaged using \verb|DPPP| and \verb|WSclean| \citep{DPPP2018,Offringa2014}. The flux density scale was set according to \cite{ScaifeHeald2012}, and was subsequently verified through comparison with the TIFR GMRT Sky Survey \citep[TGSS,][]{Intema2017}. The flux calibration error is estimated to be 10\% \citep{Hardcastle2020}.

\subsection{VLA observations and data reduction}
MS0735 has previously been observed by the VLA in a number of different bands and configurations. We used the following specific datasets:
\begin{itemize}
    \item 325 MHz, A-configuration, observed on 24 December 2004.
    \\The observation was performed using 26 antennas for a total time on target of 2.4 h, divided into four scans alternated by 2-minute observations of the phase calibrator. The flux density calibrator was observed at the end of the observation for 6 minutes. The frequency band was divided into two spectral windows of 31 channels each for a total bandwidth of 3 MHz. The data were recorded using a dual-mode polarization. 
    \item 325 MHz, B-configuration, observed on 4 December 2003.
    \\The target was observed for 3.6 h divided into three scans alternated with 3 minutes of phase calibrator observations. The flux density calibrator was observed at the beginning and end of the observing run. Twenty-six antennas were used. The observing band was divided into two sub-bands with 15 channels each of 391 kHz width for a total bandwidth of 12 MHz. 
    \item 325 MHz, C-configuration, observed on 24 December 2006.
    \\The observation was performed using 25 antennas, some equipped with the old VLA receivers, whereas the others were made with the new receivers. EVLA-VLA baselines were affected by closure errors. We corrected for this effect using the flux density calibrator. The source was observed in 14 time-scans for a total time of 4 h. The phase calibrator was observed for 2 minutes before every target scan, while the flux density calibrator was observed at the beginning and end of the observing run for 4 minutes each time. The data were recorded using two sub-bands of 31 channels each for a total bandwidth of 6 MHz.
    \item 1420 MHz, A-configuration, observed on 24 October 2004.
    \\The observation was performed using 26 antennas and two single-channel sub-bands with a width of 25 MHz each. The source was observed for roughly 2h, alternated by 2-minute observations of the phase calibrator. The flux density calibrator was instead observed at the end of the observation for 3 minutes. 
    \item 1420 MHz, B-configuration, observed on 19 August 2006.
    \\Observation performed using 25 antennas, 4 of which were equipped with the old VLA receivers. These were flagged from the dataset because they clearly provided a different response with respect to the upgraded antenna, and no suitable calibrator was observed to correct for this effect. The data were recorded in two single-channel sub-bands with a width of 50 MHz. The source was observed for a total time of 5h divided into 12-minute time-scans. The flux density calibrator was observed at the beginning and end of the observation (4 minutes each time), while the phase calibrator was observed alternating with the source for 3 minutes every time-scan.
    \item 1420 MHz, C-configuration, observed on 15 April 2004.
    \\Observation performed using 27 antennas, two sub-bands of seven channels each for a total bandwidth of 22 MHz. The data were recorded for a total integration time on target of 3.4 h divided into five time-scans. The flux density calibrator was observed at the beginning and end of the observing run for a total time of 10 minutes; the phase calibrator was observed for 2 minutes alternating with scans on target.
    \item 8460 MHz, A-configuration, observed on 17 October 2004.
    \\The source was observed for 1.7h using 25 antennas. The data were recorded in two single-channel sub-bands with a width of 25 MHz each.
    The phase calibrator was observed alternating with the target for 2 minutes each time, while the flux density calibrator was observed at the end of the observation for 2 minutes. 
    \item 8460 MHz, D-configuration, observed on 28 November 2005.
    \\ The observation was performed using 23 antennas and two single-channel spectral windows centred at 8440 MHz and 8490 MHz with a bandwidth of 50 MHz each. The target was observed for 4.5 h divided into ten scans interspersed with 2-minute phase calibrator observations. The flux density calibrator was observed for 8 minutes at the beginning of the run.
\end{itemize}

All datasets were reduced with Common Astronomy Software Applications \citep[{\tt{CASA}}, version 5.4;][]{McMullin2007} using the following steps: after manual flagging, we performed calibration in the standard manner. The flux density scale was set according to \cite{PerleyButler2013} for all datasets except for the data at 325 MHz, for which we used the scale of \cite{ScaifeHeald2012}. The flux calibration error is estimated to be 5\% for P band and L band and 2\% for X band \citep{ScaifeHeald2012, PerleyButler2013}. 

Following standard calibration, we performed several cycles of phase-only and amplitude-and-phase self-calibration at 325 MHz and 1420 MHz. Self-calibration was also attempted at 8460 MHz, but due to the low flux density of the source, the process did not improve the gain solutions and the solutions were not applied. Finally, we combined all individual datasets at matching frequencies (325 MHz, 1420 MHz, and 8460 MHz) and jointly deconvolved them to improve the \emph{uv}-coverage.

\subsection{GMRT observations and data reduction}
The source was observed with the GMRT at 235 MHz and 610 MHz on 4 December 2009 in dual-frequency mode using the GMRT Hardware Backend (GHB). In observations performed with GHB, the 32 MHz bandwidth is typically split over an upper-side band (USB) and lower-side band (LSB), both recorded in separate LTA files.

For the data at 235 MHz we were able to calibrate only the USB because good calibrator scans are lacking in the LSB. The bandwidth is thus limited to 16 MHz. The target was observed in nine time-scans for a total integration time of 6 h. The source 3C286 was used as flux-density calibrator and observed at the end of the observing run for 12 minutes. 

The flux calibration error is estimated to be 10\% \citep{Chandra2004}.
We processed the data using the Source Peeling and Atmospheric Modelling \citep[{\tt{SPAM}};][]{Intema2014, Intema2017} pipeline that corrects for ionospheric direction-dependent effects. The output calibrated visibility data were then imported into \verb|CASA| for imaging.

\subsection{X-ray observations}
We used the \emph{Chandra} data already published in \cite{Vantyghem2014} to perform some further analysis as described in Sect. \ref{sec:X}.
The source was observed seven times with \emph{Chandra} in June 2009 for a cumulative exposure time of 477 ks. Each observation was calibrated in the standard way \citep[see ][for more details]{Vantyghem2014} and was then imaged by summing events in the energy range 0.5–7.0 keV. These images were then combined to create a single mosaic image to study the X-ray emission. Point sources were identified and removed. Finally, the best-fitting double-$\beta$ model was subtracted from the cluster emission, producing the residual image we report in the left and middle panels of Fig. \ref{fig:X}.

\section{Results}\label{sec:results}

\subsection{Radio morphology}
In Fig. \ref{fig:Composite} we show a composite image of MS0735, obtained by combining the Hubble Space Telescope I-band image (yellow), the \emph{Chandra} X-ray image (blue), and the radio image at two different frequencies, LOFAR 144 MHz (orange) and VLA 325 MHz (green), at a similar resolution of 7"x5", to compare the extension of the radio emission.
The radio lobes at LOFAR frequency are clearly wider than previously found with the VLA at higher frequencies and now completely fill the X-ray cavities.
In Fig. \ref{fig:lofar} we show the high-resolution image of MS0735 as seen with the LOFAR-HBA at a central frequency of 144 MHz. The two radio lobes, called the northern and southern outer lobes, extend in the north-south direction for $\sim75\asec\cong 260$ kpc from the bright central emission. At the extreme edge of the lobes an increase in surface brightness is visible, which can be interpreted as hotspots.

\begin{figure*}
\centering
\includegraphics[width=17cm]{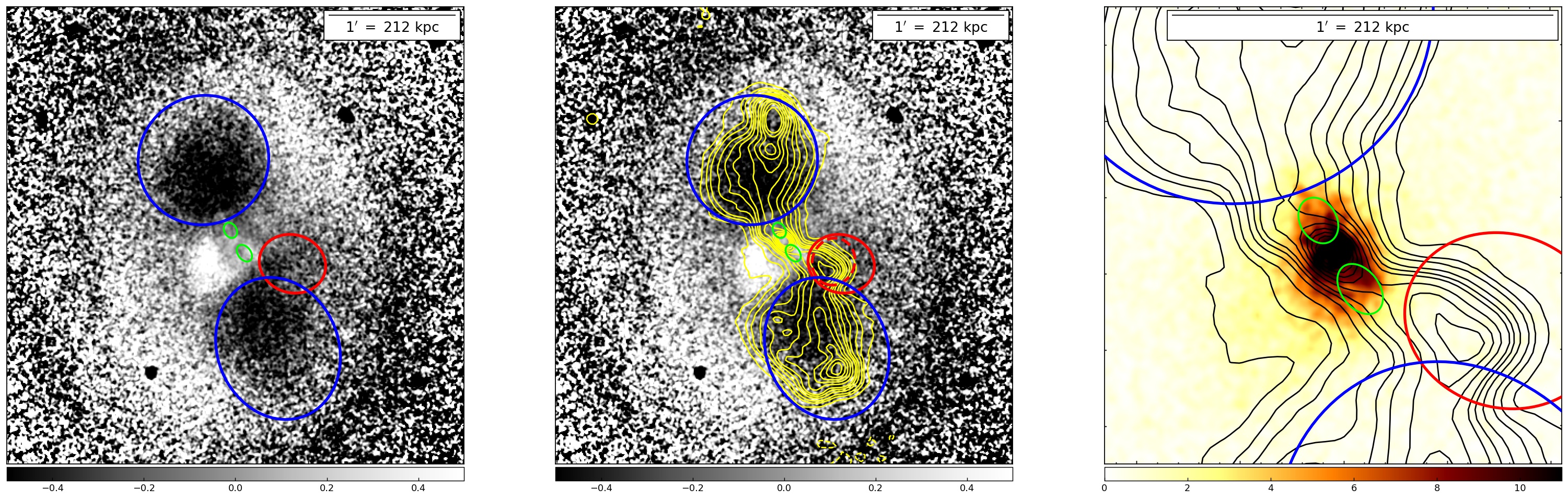}
\caption{\emph{Chandra} images of MS0735. \emph{Left:} Residual image after subtracting a double$-\beta$ model from the X-ray image (0.5$-$7 keV). The image is in units of counts $\rm pixel^{-1}$ and is Gaussian-smoothed with a 1 arcsec kernel radius. The dark regions to the north and south correspond to two large cavities, marked by blue ellipses. A surface brightness edge surrounds these cavities and corresponds to a weak shock front. The red ellipse marks the newly discovered intermediate cavity south-west of the core, while green ellipses show the inner cavities. \emph{Middle:} Same as before, with the LOFAR contours at 144 MHz superposed, and the smaller region outlined by the dashed red ellipse is considered to place a lower limit on the size of the intermediate cavity. \emph{Right:} Soft-band (0.3$-$1.0 keV) image in units of counts $\rm pixel^{-1}$, Gaussian-smoothed with a 1 arcsec kernel radius. Two regions of low surface brightness, interpreted as a pair of cavities originating from a recent AGN outburst, are indicated with green ellipses. The LOFAR contours at 144 MHz and the blue and red ellipses showing the outer and intermediate cavities, respectively, are superposed.}
\label{fig:X}
\end{figure*}

At this resolution the extended emission south-west of the core is clearly visible, which was previously noted at higher frequencies by \cite{Cohen2005}. 
This was suggested to either be a signature of a second epoch of AGN activity or a bending of the southern jet direction because of the interaction with the ICM.
Our new analysis of the X-ray data, as discussed in Sect. \ref{sec:X}, points to the presence of a further cavity corresponding to this radio emission, which was not noticed before.
Together with the radio morphology observed at high resolution (see LOFAR and VLA images, \ref{fig:lofar} and \ref{fig:all_images}), this supports the idea that this emission represents a radio lobe associated with a further phase of jet activity that occurs between the two jet activities that were previously discovered by \cite{Vantyghem2014}. 
It is visible in the central panel of Fig. \ref{fig:X}, where the LOFAR contours are superposed on the X-ray image, and the radio emission fills the newly detected cavity, represented with a red ellipse. From here on, we refer to this emission as to the intermediate lobe. At its edge seems to be a hotspot, based on the increase in surface brightness. The bending of this emission with respect to the outer lobes indicates that the reactivated jet has changed its direction. No counterpart of this intermediate lobe, north-east of the core, is detected. This could be due to projection effects, for instance, Doppler boosting \citep[e.g.][]{Blandford1979, Orr1982, Kellermann1988, Hardcastle1998}.

From \cite{Vantyghem2014} we know that a third series of cavities is present in the innermost regions of the source. These are shown in the right panel of Fig. \ref{fig:X} with green ellipses, where the soft band X-ray image is superposed with the LOFAR contours. The limited resolution of the radio images does not allow us to detect the corresponding inner lobes or jets, however.

The central panel of Fig. \ref{fig:X} also shows that all the radio emission is confined to the surrounding cocoon. The high energy associated with the outburst has displaced and compressed the material in the cluster atmosphere, creating a confining shell that hinders the advance of any radio plasma.

We also checked for more diffuse emission by re-imaging the LOFAR data at lower resolution. The overall morphology of the source in this case is consistent with what was previously observed at 325 MHz \citep{Cohen2005}. While the radio lobes are more extended at 144 MHz, we find no evidence of any older outbursts. 

In Fig. \ref{fig:all_images} we show images of MS0735 at different frequencies, made from the archival data described earlier.
At higher frequencies, the morphology of the source changes: the central region (which includes the core and the inner jets) becomes dominant, while the lobes become less prominent, until they are no longer visible at 8460 MHz. To verify that this variation is real or due to different observational setups, we re-imaged all the datasets using the same \emph{uv}-range, with a \verb|uniform| weighting scheme, and convolved all images to the same resolution (see the low-resolution column of Table \ref{tab:image_parameter}).
The morphological differences remain even in this case, indicating that the lack of lobe emission at 8460 MHz is really due to the spectral steepening and not to instrumental effects.
This implies a sharp cut-off in the spectrum of the outer lobes that is typical of aged plasma.

When we consider the morphology at different frequencies, 
MS0735 may overcome the boundaries of the Fanaroff-Riley (FR) classification \citep{FR1974} in that it could be at different points of the activity cycle.
The outer lobes exhibit the classical morphology of FRII radio galaxies, with lobes and hotspots (visible also in the VLA and GMRT images up to 1420 MHz). The second phase of activity may have created another FRII, presenting an increase in brightness in the intermediate lobe, but at 610 MHz, the jets are clearly visible, which is unusual for this class of source.
It is therefore possible that the source was born as an FRII, and was then slowed down by the dense surrounding medium at the centre of the cool-core cluster. In this scenario, the source could have turned into an FRI-type radio galaxy during subsequent phases of jet activity \citep[e.g.][]{Laing1994,KaiserBest2007,Meliani2008,Wang2011,TurnerShabala2015,Kapinska2017}. 

Understanding the nature of the source and its variation throughout its life is beyond the scope of this paper. For our purposes, that is, estimating the radiative life of the outer lobes and of the intermediate lobe, we assume that the source is an FRII, so the particle acceleration took place in the hotspots.
In the following sections, we discuss this point further.

\begin{table*}
    \centering
    \caption{Cavity properties.}
    \renewcommand\arraystretch{1.2}
    \begin{tabular}{ccccccccc} \hline
    Cavity &a &b &R &$E_{\rm cav}$ &$t_{\rm buoy}$ &$t_{\rm c_{s}}$ &$t_{\rm ref}$  &$P_{\rm cav,buoy}$\\
    &[kpc] &[kpc] &[kpc] &[$10^{59}$ erg] &[Myr] &[Myr] &[Myr] &[$10^{44}\ \rm{erg\ s^{-1}}$]\\\hline
Northern outer  &$109\pm16$ &$106\pm16$ &150 &$440\pm200$ &91 &120 &240 &$160\pm70$\\
Southern outer &$120\pm15$ &$100\pm21$ &186 &$440\pm200$ &110 &140 &250 &$130\pm60$\\
Intermediate  &$48\pm9$ &$42\pm7$ &85 - 100 &23 - 60 &62 - 64 &53 - 69 &103 - 120  &12 - 30\\
Northern inner &$13.3\pm1.6$ &$10.1\pm2.3$ &19.3 &$3.6\pm2.0$ &33 &19 &74 &$3.6\pm2.0$\\
Southern inner  &$15.5\pm2.2$ &$10.5\pm3.3$ &25.0 &$3.6\pm2.4$ &41 &25 &78 &$2.9\pm2.0$\\\hline
    \end{tabular}
    \tablefoot{For each cavity we list the semi-minor axis ($a$), semi-major axis ($b$), distance from cluster centre ($R$), energy of the cavity ($E_{\rm cav}$), buoyancy timescale ($t_{\rm buoy}$), sound crossing timescale ($t_{\rm c_s}$) refilling timescale ($t_{\rm ref}$), and power of the cavity ($P_{\rm cav, buoy} = E_{\rm cav}/t_{\rm buoy}$). The values for the outer and inner cavities are taken from Table 2 of \cite{Vantyghem2014}.}
    \label{tab:X_cavity}
\end{table*}

\subsection{X-ray features}\label{sec:X}
\cite{Vantyghem2014} detected two large cavities, with a diameter of $\sim200$ kpc and a shock front enveloping the cavities. They are both clearly visible in the left and middle panels of Fig. \ref{fig:X}.
The projected sizes and positions of the cavities were determined
by eye by approximating the surface brightness depressions with elliptical regions, see the blue ellipses in Fig. \ref{fig:X}.
The northern cavity is best represented by an ellipse with a semi-major axis a = $109\pm16$ kpc, semi-minor axis b = $106\pm16$ kpc, and projected distance from the AGN, R = 150 kpc. The southern cavity is best represented by an ellipse with a = $120\pm15$, b = $100\pm21$, and R = 186 kpc.
The cocoon shock front is approximated by an ellipse with a semi-major axis a = 320 kpc and a semi-minor axis b = 230 kpc.

The deep \emph{Chandra} image in the soft band (0.3 - 1.0 keV), right panel of Fig. \ref{fig:X}, revealed two smaller cavities located along the radio jets in the inner 20 kpc. Their positions and sizes are shown in all the panels of Fig. \ref{fig:X} with green ellipses. 
These cavities are best represented by an ellipse with a = $13.3\pm1.6$ kpc, b = $10.1\pm2.3$ kpc, and a projected distance of R = 19.3 kpc from the cluster centre for the inner northern cavity and a = $15.5\pm2.2$, b = $10.5\pm3.3,$ and R = 25.0 kpc for the inner southern cavity.

We performed a further visual inspection of the X-ray residual image that led to the possible detection of an additional cavity, located to the south-west of the core at an intermediate distance from the cluster centre with respect to that of the two cavities reported in \cite{Vantyghem2014}. No corresponding cavity to the north-east of the core has been detected.
This newly identified cavity (referred to as the intermediate cavity), represented with a red ellipse in Fig. \ref{fig:X}, is distinct from the region used for the analysis of the outer cavity to the south \citep{Vantyghem2014}, represented with a blue ellipse.
Because the radio emission is less sensitive to projection effects than depressions in the X-ray image, we considered different dimensions for the intermediate cavity \citep[as done, e.g. in][]{Gitti2010}, ranging from the size of the apparent depression seen in the residual X-ray image (solid red ellipse) to the extension of the intermediate radio lobe (dashed red ellipse in middle panel of Fig. \ref{fig:X}). Our analysis provides a range of values for a semi-major axis a = 39 - 56 kpc, semi-minor axis b = 35 - 48 kpc, and distance from the cluster centre R = 85 - 100 kpc.

We estimated the energetics and age of this newly detected intermediate cavity following the same procedure as was used by \citet{Vantyghem2014} for the other cavities. 
The minimum total energy required to create a cavity for a relativistic gas is $E_{\rm cav} = 4pV$, where $p$ is the cavity pressure and V is its volume. Assuming pressure equilibrium, the pressure is estimated from the deprojected pressure profile of the surrounding ICM \citep[Fig. 7 of][]{Vantyghem2014} at a radius corresponding to the distance of the cavity centre. The cavity volume is calculated using the geometric mean between oblate and prolate ellipsoids, $V = 4/3\pi (ab)^{3/2}$. The age of the cavity is estimated using three
characteristic timescales: the sound crossing time, buoyancy time,
and refill time \citep[e.g.][]{Birzan2004}. 
At the distance of the intermediate cavity, we estimated a sound velocity of $c_s=1210-1260$ km s$^{-1}$.
The gravitational acceleration, g, used to calculate the buoyancy and refill timescales was determined using the MS0735 mass profile from \cite{Hogan2017}.
Finally, we estimated the power required to inflate the cavity, as $P_{\rm cav} = 4pV/t_{\rm buoy}$, where we used the buoyancy timescale as reference for the cavity age.
The properties and energetics of the intermediate cavity derived in this work, along with those of the inner and outer cavities studied by \citet{Vantyghem2014}, are reported in Table \ref{tab:X_cavity}.

The X-ray data have thus revealed that there were at least three AGN outbursts in MS0735. Only few sources with multiple cavities are currently known, for example Hydra A \citep{Nulsen2005,Wise2007,Gitti2011} and NGC 5813 \citep{Randall2011,Randall2015}.
The presence of different episodes of jet activity is important to ensure the continuous heating of the central gas and so to prevent it from cooling.

\subsection{Radio spectral study}
To perform a spectral study of the source, we re-imaged all the radio data with the same restoring beam ($16\asec\times10\asec$, PA $0^{\circ}$), \emph{uv}-range ($157\sim19173\ \lambda$), and used a \verb|uniform| weighting scheme. 
We excluded the data at 8460 MHz, where the emission of the lobes is not visible.
Finally, we spatially aligned all the images to correct for possible shifts introduced by the phase self-calibration process, using only the central compact emission as reference because no point sources are visible in the vicinity of the target. After this procedure, the spatial difference between the images was reduced to less than 0.1 pixel.

For a more detailed study of the central regions of the source, we furthermore created another set of images at higher resolution, excluding the GMRT data at 235 MHz. The exclusion of this frequency has very little effect on the results of the study because it lies between two points at frequencies very close to each other. These high-resolution images have a restoring beam of $6.5\asec$.
The parameters of the two sets of images are listed in Table \ref{tab:image_parameter}.

\begin{figure}
    \resizebox{\hsize}{!}{\includegraphics{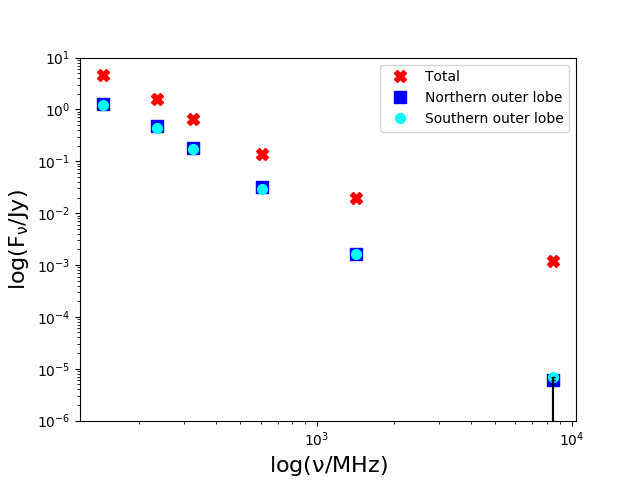}}
    \caption{\footnotesize{Total flux density distribution of MS0735 between 144 MHz and 1420 MHz (in red) and of the northern and southern outer lobes (in blue and cyan, respectively). We have also set a $3\sigma$ upper limit for the flux of the lobes at 8460 MHz because they are not detected.}}
    \label{fig:Tot_flux}
\end{figure}

From the images at lower resolution, we measured the total flux density of the source and of the outer lobes, reported in Table \ref{tab:flux} and plotted in Fig. \ref{fig:Tot_flux}.
We also set an upper limit for the lobes emission at 8460 MHz, estimated as the product between the rms noise at 8460 MHz and the lobe area expressed in units of beams ($3\sigma\times N_{\rm beam})$. 

We used the Broadband Radio Astronomy Tools (\verb|BRATS|)\footnote{http://www.askanastronomer.co.uk/brats/} software package \citep{Harwood2013, Harwood2015} to estimate the spectral age of the source. BRATS allows performing a detailed spectral analysis on a pixel-by-pixel basis. In this way, we mapped the source properties (e.g. spectral index, spectral age) throughout the source extension. The detection limit was set to three times the noise, estimated as the rms in an empty region in the field.

\begin{table}
    \centering
     \renewcommand\arraystretch{1.2}
    \caption{Summary of imaging parameters.}
    \begin{tabular}{ccc}\hline
        Parameter &High resolution & Low resolution  \\\hline
        Frequency range & 144 - 1420 MHz & 144 - 1420 MHz \\
        & (excluding 235 MHz) & (all) \\
        Restoring beam & $6.5\asec \times 6.5\asec$ & $16\asec \times 10\asec$, PA $0^{\circ}$ \\
        \emph{uv}-range & $157\sim19173\ \lambda$ & $157\sim40186\ \lambda$ \\
        Weighting & {\tt{uniform}} & {\tt{uniform}} \\\hline
    \end{tabular}
    \label{tab:image_parameter}
\end{table}

\begin{table*}
    \centering
    \caption{Total flux density of MS0735 and of the outer lobes.}
     \renewcommand\arraystretch{1.2}
    \begin{tabular}{ccccc} \hline
         Telescope &$\nu$ &Total &Northern outer lobe &Southern outer lobe \\
         &[MHz] &[Jy] &[Jy] &[Jy] \\\hline
         LOFAR &144 &$4.7\pm0.5$ &$1.3\pm0.1$ &$1.2\pm0.1$\\
         GMRT &235 &$1.6\pm0.2$ &$(4.8\pm0.5)\times10^{-1}$ &$(4.4\pm0.4)\times10^{-1}$ \\
         VLA &325 &$(6.5\pm0.3)\times10^{-1}$ &$(1.8\pm0.1)\times10^{-1}$ &$(1.8\pm0.1)\times10^{-1}$\\
         GMRT &610 &$(1.4\pm0.1)\times10^{-1}$ &$(3.1\pm0.3)\times10^{-2}$ &$(3.0\pm0.3)\times10^{-2}$ \\
         VLA &1420 &$(2.0\pm0.2)\times10^{-2}$ &$(1.7\pm0.1)\times10^{-3}$ &$(1.6\pm0.1)\times10^{-3}$\\
         VLA &8460 &$(1.20\pm0.03)\times10^{-3}$ &$(\sim6.1\times10^{-6})$ &($\sim6.9\times10^{-6}$)\\\hline
    \end{tabular}
    \tablefoot{$3\sigma$ upper limit are shown at 8460 MHz for the lobes flux density because they are not detected.}
    \label{tab:flux}
\end{table*}

\begin{figure*}
\centering
    \includegraphics[width=17cm]{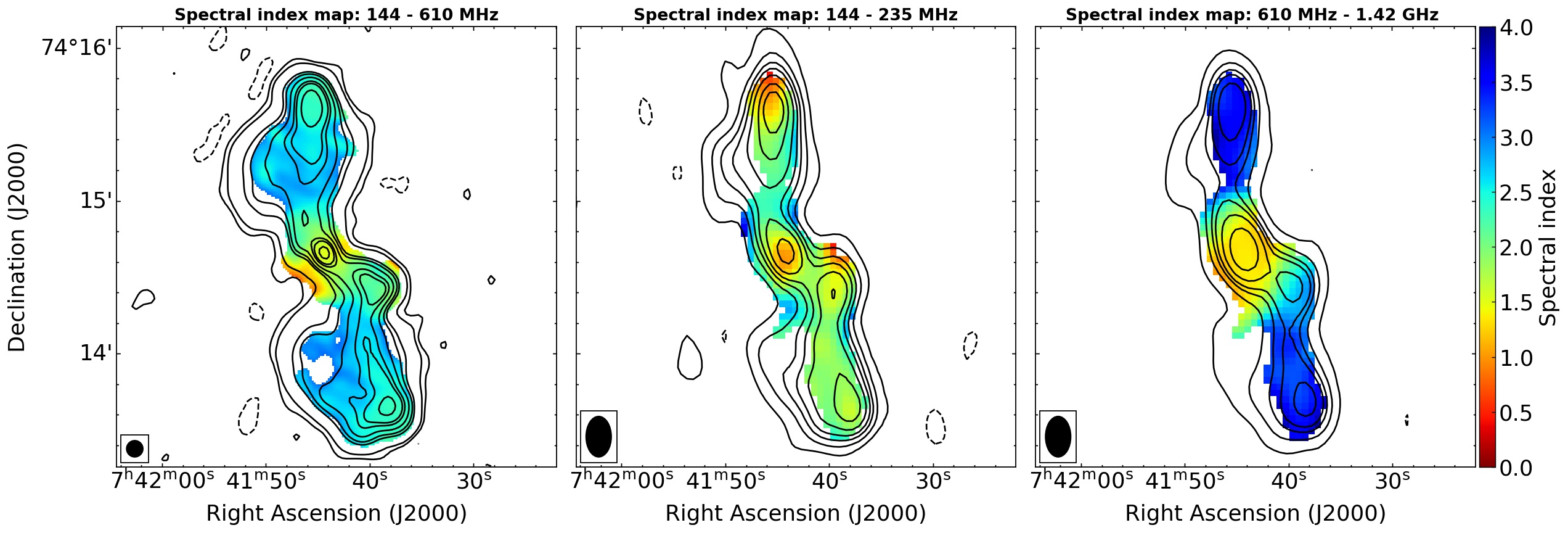}
    \caption{Spectral index maps of MS0735.\emph{Left:} High-resolution spectral index map between 144 MHz and 610 MHz, resolution = $6.5\asec$. The LOFAR contours are overlaid. \emph{Middle:} Low-frequency spectral index map between 144 MHz and 235 MHz, resolution = $16\asec\times10\asec$. The 235 MHz contours are overlaid. \emph{Right:} High-frequency spectral index map between 610 MHz and 1420 MHz, resolution = $16\asec\times10\asec$. The 610 MHz contours are overlaid. The beam is shown in the bottom left corner of each map.}
    \label{fig:spix}
\end{figure*}

\subsubsection{Spectral index maps}
To investigate the spatial distribution of the spectral index, we made a high-resolution ($6.5\asec$) two-frequency spectral index map using LOFAR 144 MHz and GMRT 610 MHz images (see the left panel of Fig. \ref{fig:spix}).
To our knowledge, this is the first time that a spectral index study of MS0735 extended to this low frequency is presented at high resolution.

Previous work that was based on integrated spectra found an ultra-steep radio spectrum for this source ($\alpha_{325}^{1425}=2.45\pm0.04$ \citeauthor{Cohen2005} \citeyear{Cohen2005}, $\alpha_{327}^{1400}=2.48\pm0.04$ \citeauthor{Birzan2008} \citeyear{Birzan2008}). 
Our resolved spectral study confirms that this trend extends to 144 MHz, with spectral index values reaching $\alpha_{144}^{610}=3.1\pm0.3$. Curiously, the spectrum remains very steep even at the location of the core, where the spectral index is $\alpha_{144}^{610}=1.5\pm0.1$, as previously observed at higher frequency ($\alpha_{325}^{1425}=1.54\pm0.04$, \citeauthor{Cohen2005} \citeyear{Cohen2005}).
This steep value suggests superposition of the core and the lobe emission.
The emission from the central region is further analysed in the next section.

In addition, we note a gradient of the spectral index along the main axis of the source. Starting from a value of $\alpha_{144}^{610}=2.3\pm0.1$ in the hotspots, at the edges of the outer lobes, it steepens when the core is approached, reaching a maximum value of $\alpha_{144}^{610}=2.9\pm0.3$ in the outer lobes, then flattens again in the central regions and in the intermediate lobe, where the spectral index is $\alpha_{144}^{610}=2.1\pm0.1$.
This trend suggests two different episodes of jet activity.
If there were only one phase of jet activity, we would expect a gradual increase in the spectral index from the hotspot, where the particles are last accelerated, to the core where the oldest particles are found. Instead, the spectral index in the central regions is flatter than in the outer lobes, indicating the presence of younger particles, which were therefore emitted later.
These ultra-steep spectral indices in hotspots occur in $\leq 0.5\%$ of cases (Hogan et al. 2015). This likely indicates that the acceleration of particles is no longer active in these regions, otherwise we would expect a spectral index in the range 0.6-0.8, as observed for typical FRII-type radio galaxies \citep{JaffePerola1973,Carilli1991,Komissarov1994}.

In the middle and right panels of Fig. \ref{fig:spix}, we show the low-frequency spectral index map between 144 MHz and 235 MHz and the high-frequency spectral index map between 610 MHz and 1420 MHz, respectively, both at the resolution of $16\asec\times10\asec$. 
When the frequency range between 144 and 235 MHz is narrowed, the spectrum is less steep than that represented in the map obtained between 144 and 610 MHz. This is especially true in the hotspot of the northern outer lobe where the spectral index reaches the minimum value of $\alpha_{144}^{235}=0.75\pm0.4$, much closer to typical values.
The hotspot in the southern outer lobe is instead less pronounced and has a steeper spectral index, $\alpha_{144}^{235}=1.6\pm0.4$. This difference may be due to projection effects in the southern outer lobe, with an overlap along the line of sight of particles of different ages, making the spectrum steeper.
At higher frequencies, the spectrum is instead very steep, and we recover the values found by \cite{Cohen2005} for the centre and the outer lobes.
For the main features of the source, we list in Table \ref{tab:spix} the average spectral index at low and high frequencies and their difference, which represents the spectral curvature of these regions (SPC = $\alpha_{\rm{high}} - \alpha_{\rm{low}}$).
For the two outer lobes and the intermediate lobe, the high-frequency spectral index is steeper than that observed at low frequency, and despite the large uncertainties, the spectral curvature is high, indicating emission from an aged population of electrons.

\begin{table}
    \centering
    \renewcommand\arraystretch{1.2}
        \caption{Spectral indices ($\alpha_{\rm{low}}$, $\alpha_{\rm{high}}$) and spectral curvature (SPC) for the main features of MS0735.}
    \begin{tabular}{cccc}\hline
         Region &$\alpha_{\rm low}$ &$\alpha_{\rm high}$ & SPC \\\hline
         Centre &1.2$\pm$0.4 &1.4$\pm$0.1 & 0.2$\pm$0.5 \\
         Northern outer lobe &2.1$\pm$0.4 &3.4$\pm$0.6 &1.3$\pm$0.9\\
         Northern hotspot &0.75$\pm$0.4 &3.4$\pm$0.6 &2.6$\pm$0.9 \\
         Southern outer lobe &2.0$\pm$0.4  &3.4$\pm$0.6 &1.4$\pm$0.9\\
         Southern hotspot &1.6$\pm$0.4 &3.4$\pm$0.6 &1.8$\pm$0.9 \\
         Intermediate lobe &1.5$\pm$0.4 &2.8$\pm$0.3 &1.3$\pm$0.7\\\hline
    \end{tabular}
    \label{tab:spix}
\end{table}

\begin{figure*}
    \centering
    \includegraphics[width=17cm]{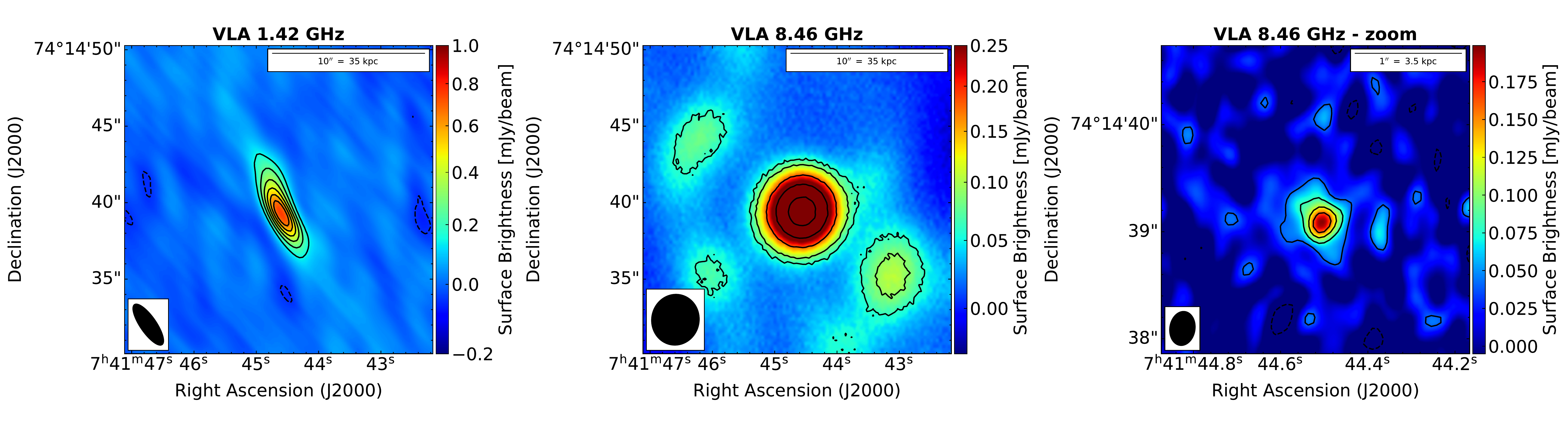}
    \caption{Radio maps of the central emission of MS0735. \emph{Left}: VLA 1420 MHz at $3.1\asec \times 1.1\asec$. Levels: [$-1$, 3, 5, 7, 9, 11, 13] $\times 3\sigma$ ($\sigma = 17\ \upmu\rm{Jy\ beam^{-1}}$). \emph{Middle}: VLA 8460 MHz at $3.3\asec \times 3.1\asec$  resolution. Levels: [$-1$, 1, 1.5, 3, 5, 8] $\times 3\sigma$ ($\sigma = 20\ \upmu\rm{Jy\ beam^{-1}}$). \emph{Right}: Zoom of VLA 8460 MHz at $0.32\asec \times 0.24\asec$ resolution. Levels: [$-1$, 1, 2, 3, 4] $\times 3\sigma$ ($\sigma = 13\ \upmu\rm{Jy\ beam^{-1}}$). The beam is shown in the bottom left corner of each image.}
    \label{fig:core}
\end{figure*}

\subsubsection{Central emission}
As explained above, the central region of MS0735 exhibits a steep radio spectrum, with $\alpha=1.5\pm0.1$ between 144$-$610 MHz, which is much steeper than that of typical AGN cores. However, we note that at the resolution of $6.5\asec\cong 23$ kpc, the actual core is likely not resolved, and the observed emission (and thus apparent steep spectrum) could result from the superposition of different populations of electrons from the inner jets or lobes. 
The detection of inner cavities in fact demonstrates that there has been more recent jet activity \citep{Vantyghem2014}.

To investigate this region further, we analysed high-resolution archival VLA observations of the source. These observations (at 1420 MHz for array A and 8460 MHz for array A, listed in Sect. 2.2) were performed without the use of short baselines, so the diffuse emission is filtered out, allowing us to study the real compact emission of the nuclear region.

In Fig. \ref{fig:core} we show the radio map of the central emission at 1420 MHz (left panel), at $3.1\asec \times 1.1\asec$ resolution, at 8460 MHz (middle panel) $3.3\asec \times 3.1\asec$ resolution, and the emission at 8460 MHz (right panel), at $0.32\asec \times 0.24\asec$ resolution.
To measure the flux densities of the central emission component at 1420 MHz and 8460 MHz, we used images produced with a common uv-range ($17926 \sim 353657\ \lambda$) and a resolution of $2\asec \cong 7$ kpc.

We obtain a spectral index of $0.75\pm0.08$ for this central component. This supports the idea that even at this resolution, the emission we observe is not purely core emission, which would have a much flatter spectrum \citep[e.g.][]{Blandford1979,Hogan2015}, but rather a superposition of emission from the core and the inner part of the jets. 

\begin{figure*}
    \centering
    \includegraphics[width=17cm]{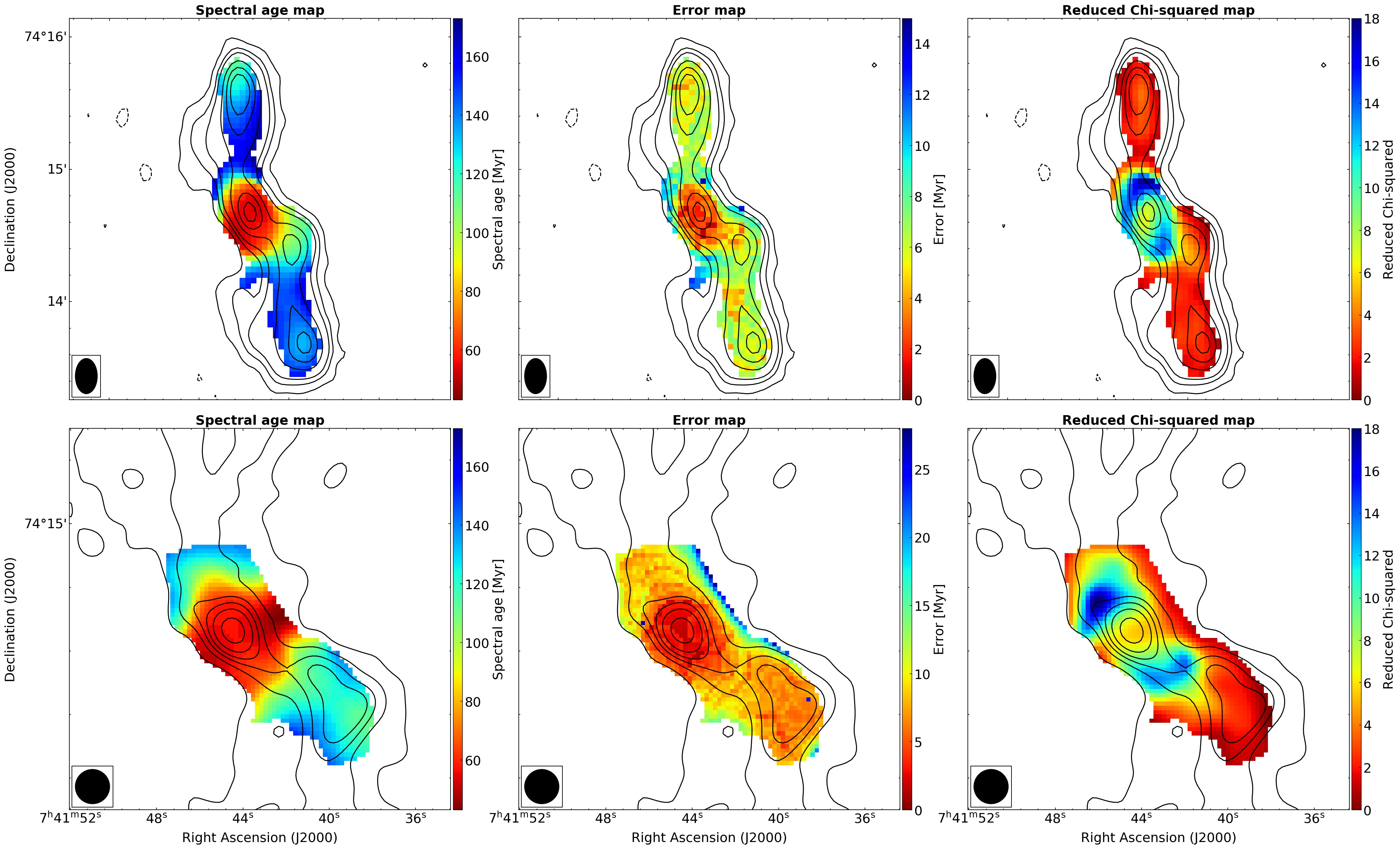}
    \caption{Tribble spectral ageing maps (\emph{left}) and corresponding error maps (\emph{middle}) and reduced chi-squared maps (\emph{right}) of the entire source at a resolution of $16\asec\times10\asec$ (\emph{top panels}) and of the central regions at a higher resolution of $6.5\asec$ (\emph{bottom panels}). The 325 MHz contours are overlaid. For the fit we set $\alpha_{\rm inj}=0.75$ and $B_{\rm eq}=5.9\ \upmu\rm{G}$.}
    \label{fig:Tribble}
\end{figure*}

\subsubsection{Radiative age}\label{sec:radiative_age}
For a synchrotron-emitting electron population, initially distributed along a power law, the energy-loss rate scales as $E/(dE/dt)\propto 1/E$. This leads to a faster cooling of high-energy electrons, and in absence of further particle acceleration, produces a spectrum that is initially described by a power law and becomes increasingly curved over time. A break in the spectrum develops at frequency $\nu_{\rm b}$, which relates to the time elapsed since the injection and to the magnetic field as $\nu_{\rm b}\propto B^{-3}t^{-2}$. This means that if we know the magnetic field strength, we can derive the age of a radio source from the shape of its spectrum.

A modern approach to deriving the spectral age of the source consists of fitting the observed radio spectrum with a modelled spectrum that is obtained by numerical integration of the equations that describe the radiative losses of the plasma.
\verb|BRATS| follows this approach and allows fitting the observed radio spectrum with different spectral ageing models, obtaining a spectral age map of the source. On small scales, particles can likely be assumed to be part of a single injection event. 

Three single-injection models are usually considered: the JP \citep{JaffePerola1973}, KP \citep{Kardashev1962, Pacholczyk1970} and Tribble \citep{Tribble1993} models. The JP and KP models consider synchrotron and inverse-Compton losses in a constant magnetic field environment, where the radio-emitting electron population has a fixed pitch angle (in the KP model) or a continuously isotropized pitch angle distribution (in the JP model). The JP model is more realistic from a physical point of view because an anisotropic pitch angle distribution will become more isotropic due to scattering. 

The Tribble model is a more complex model that attempts to account for a spatially non-uniform magnetic field by integrating the standard JP losses over a Maxwell–Boltzmann magnetic field distribution. The model implemented in the \verb|BRATS| code is a specific version that is the low-field, high-diffusion case \citep{Harwood2013, Hardcastle2013}.

Major differences between the models are visible at frequencies higher than the break frequency ($\nu_{\rm b}$), while at lower frequencies all models are expected to have a spectral index equal to the injection index ($\alpha_{\rm inj}$), which describes the initial distribution of the electron population.
We tested all the models and found that the final numbers are consistent within $10\%$. 
Therefore we report  for simplicity only the results obtained by fitting the Tribble model, which, as tested on other sources in the literature \citep[e.g.][]{Harwood2013}, provides both a good fit to observations and a more physically realistic description of the source. 

We fit the models over a range of injection indices and found that the lowest chi-squared value is obtained with $\alpha_{\rm inj}=1.5$. This injection index is much steeper than classical values
\citep[between 0.5 and 0.7, see e.g.][]{JaffePerola1973,Carilli1991, Komissarov1994} and cannot physically be explained with current particle acceleration models.
It is also steeper than the highest value found up to now, 1.2 for the cluster-centre radio galaxy 3C28 \citep[][]{Harwood2015}, whose physical interpretation has been attributed to the jet terminating in a weak shock.
However, the injection index can also be constrained by the low-frequency spectrum of the regions where particles are accelerated, that is, in the hotspots or jets if the source is an FRII or FRI, respectively.
The MHz-frequency part of the spectrum is the least affected by the ageing and can preserve information on the shape of the original energy distribution of the injected particles.
We therefore used an injection index of 0.75 for our modelling, which is the lowest spectral index value found in the hotspot of the northern outer lobe in the frequency range 144-325 MHz and is also consistent with values from the literature that were recently found from resolved spectral studies of FRII radio galaxies \citep[0.75$-$0.85; ][]{Harwood2013,Harwood2015,Harwood2017b,Shulevski2017}.
For completeness, we report the results obtained
with the well-fitting injection index value of 1.5 in Appendix \ref{sec:appendix}.

To derive the magnetic field in the outer lobes, we assumed equipartition between particles and magnetic field.
To perform this analysis we assumed a power-law particle distribution of the form $N(\gamma)\propto\gamma^{-\delta}$ between a minimum and maximum Lorentz factor of $\gamma_{\rm min} = 100$ and $\gamma_{\rm max}= 10^7$ \citep{Falcke1995, Reynolds1996, Dunn2006}.
The particle energy power-law index $(\delta)$ is related to the injection index by $\delta = 2\alpha_{\rm inj}+1$.
We also assumed that the particle energy was equally divided between electrons and protons, setting  $k = E_{\rm p}/E_{\rm e} = 1$.
Finally, we derived the flux density and the volume of the lobes at the LOFAR frequency, approximating the lobe geometry to a prolate ellipsoid with $V = (4\pi/3)ab^2$, where $a$ and $b$ are the semi-major and semi-minor axes, respectively. For the northern lobe $a=31.5\asec$ and $b=24.6\asec$, while for the southern lobe $a=27\asec$ and $b=26\asec$. The total flux of the lobes at 144 MHz is 3.3 Jy.
For $\alpha_{\rm inj}=0.75$, we derive a magnetic field strength of $B_{\rm eq}\sim 5.9\ \upmu\rm{G}$, in agreement with the equipartition magnetic field of $4.7\ \upmu$G derived by \cite{Birzan2008}.

\begin{figure}
    \resizebox{\hsize}{!}{\includegraphics{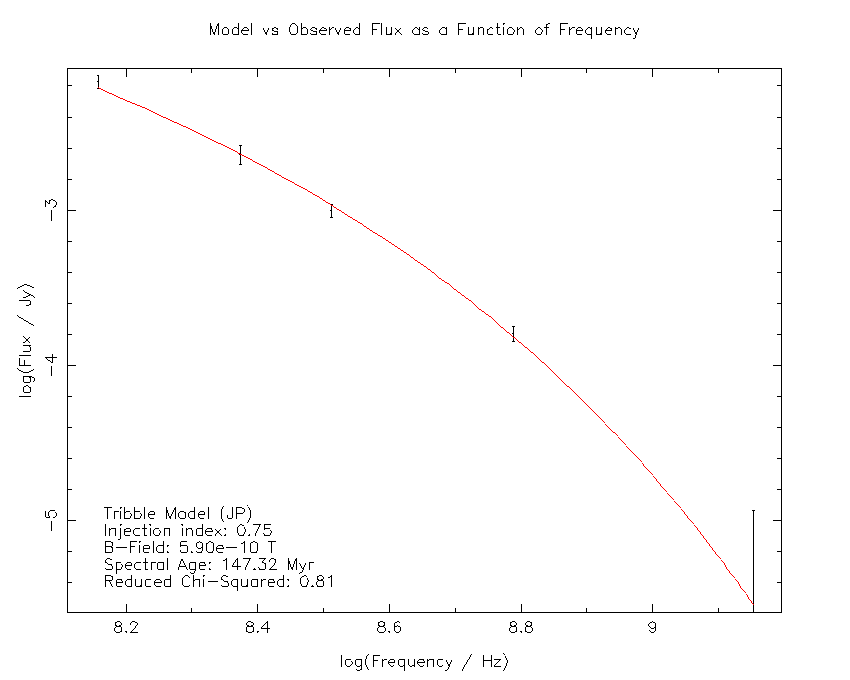}}
    \caption{Flux density distribution of an example region in the southern outer lobe, fitted with the Tribble model, setting $\alpha_{\rm inj}=0.75$ and $B_{\rm eq}=5.9\ \upmu\rm{G}$.}
    \label{fig:reg_Tribble}
\end{figure}

To derive the age of the intermediate lobe, we used the values of the injection index and magnetic field derived for the outer lobes, and we repeated the fit on the high-resolution images of the source.
In Fig. \ref{fig:Tribble} we show the spectral age map, error map, and chi-squared map of the whole source (\emph{upper panels}) and of the high-resolution zoom on the central regions (\emph{bottom panels}).
The results of the fit are reported in Table \ref{tab:CIoff}, while in Fig. \ref{fig:reg_Tribble} we show one representative spectral plot of a well-fitted single region in the southern outer lobe for the purpose of illustration.

We can reconstruct the first phase of jet activity from the ages obtained in the outer lobes. 
The oldest age found in the lobes represents the total age of the source, while the youngest age represents the OFF time, that is, the time elapsed since the last particle acceleration in the outer lobes. Consequently, we can estimate the duration of the first active phase (ON time) as the difference between the oldest and the youngest age found in the outer lobes: $t_{\rm ON} = t_ {\rm max} - t_{\rm min}$, following the approach in \cite{Shulevski2017} and \cite{Brienza2020}. The total age, OFF age, and ON age for the northern and southern outer lobes are reported in Table \ref{tab:CIoff}.

To estimate the spectral age of the intermediate lobe, we refer to the high-resolution spectral age map shown in the bottom panels of Fig. \ref{fig:Tribble}. However, this region shows only a slight age gradient from the hotspot towards the centre that is not significant enough to obtain robust values for the oldest and youngest age, due to both resolution and the presence of different components (southern outer lobe, intermediate lobe, and central emission).
Therefore we can only give an estimate of the duration of the second phase of jet activity and cannot distinguish between the active and off time. 

\vspace{0.2in}
To complete the spectral age study of MS0735, we also considered a second category of models that assumed a continuous injection of particles: the continuous injection (CI), and the so-called CIoff models \citep{Pacholczyk1970, Komissarov1994}. 
With respect to single-injection models, the source in the CI models is fuelled at a constant rate by the nuclear activity for a duration $t_{\rm ON}$ (the 'continuous injection phase'). In this phase the radio spectrum presents a first break frequency that depends on the total age of the source $t_{\rm tot}$:
\begin{equation}
    \nu^{'}_{\rm break} \propto \frac{B}{(B^2+B^2_{\rm CMB})^2 t^2_{\rm tot}}
,\end{equation}
where $B_{\rm CMB} = 3.18(1+z)^2$ is the inverse-Compton equivalent magnetic field. Below and above $\nu^{'}_{\rm break}$ the spectral indices are $\alpha_{\rm inj}$ and $\alpha_{\rm inj}+0.5$, respectively.

While in the CI model the source is continuously fuelled by fresh particles, in the CIoff, at the time $t_{\rm ON}$, the power supply of the nucleus is switched off and a new phase of duration $t_{\rm OFF}$ begins (the 'dying phase'). A new break frequency then appears beyond which the radiation spectrum drops precipitously,
\begin{equation}
    \nu^{''}_{\rm break} = \nu^{'}_{\rm break} \biggl(\frac{t_{\rm tot}}{t_{\rm OFF}}\biggl)^2
,\end{equation}
with $t_{\rm tot} = t_{\rm ON}+t_{\rm OFF}$. Thus, the CIoff model is described by four parameters: the injection index, the two break frequencies, and the normalization. As in the JP model, the synchrotron and inverse-Compton losses are modelled using a continuously isotropized pitch angle.

\begin{figure}
    \resizebox{\hsize}{!}{\includegraphics{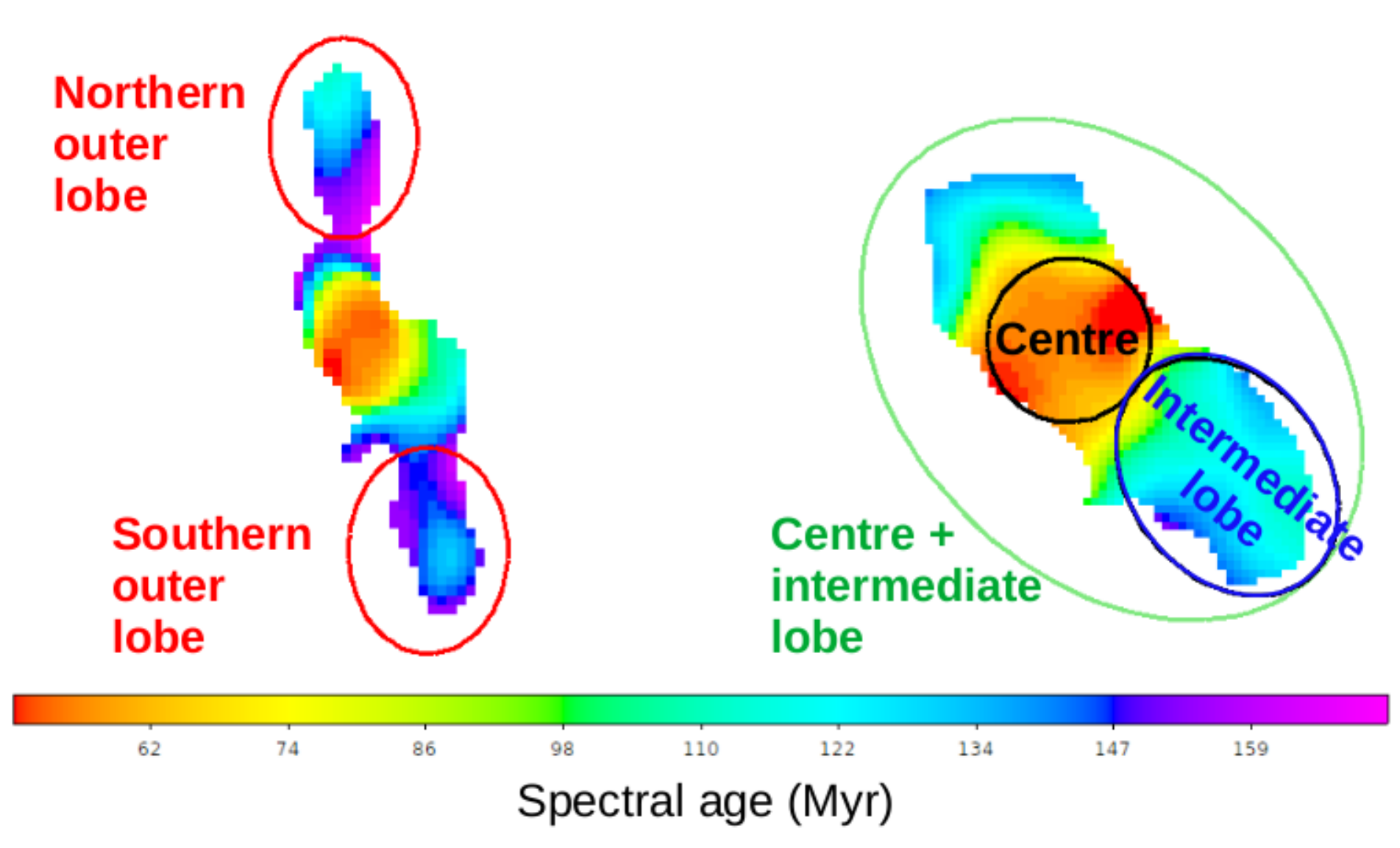}}
    \caption{Representation of the regions used to measure the integrated flux density of the main features of the source, selected based on their age. The flux density of the outer lobes is measured from the low resolution images, while we used the higher resolution images to measure the flux of the centre and of the intermediate lobe.}
    \label{fig:Regioni}
\end{figure}

These models are applicable only if the injected particles are confined to the fitted regions, so we can fit them only to the integrated flux of selected regions, shown in Fig. \ref{fig:Regioni}.

\begin{figure}
    \resizebox{\hsize}{!}{\includegraphics{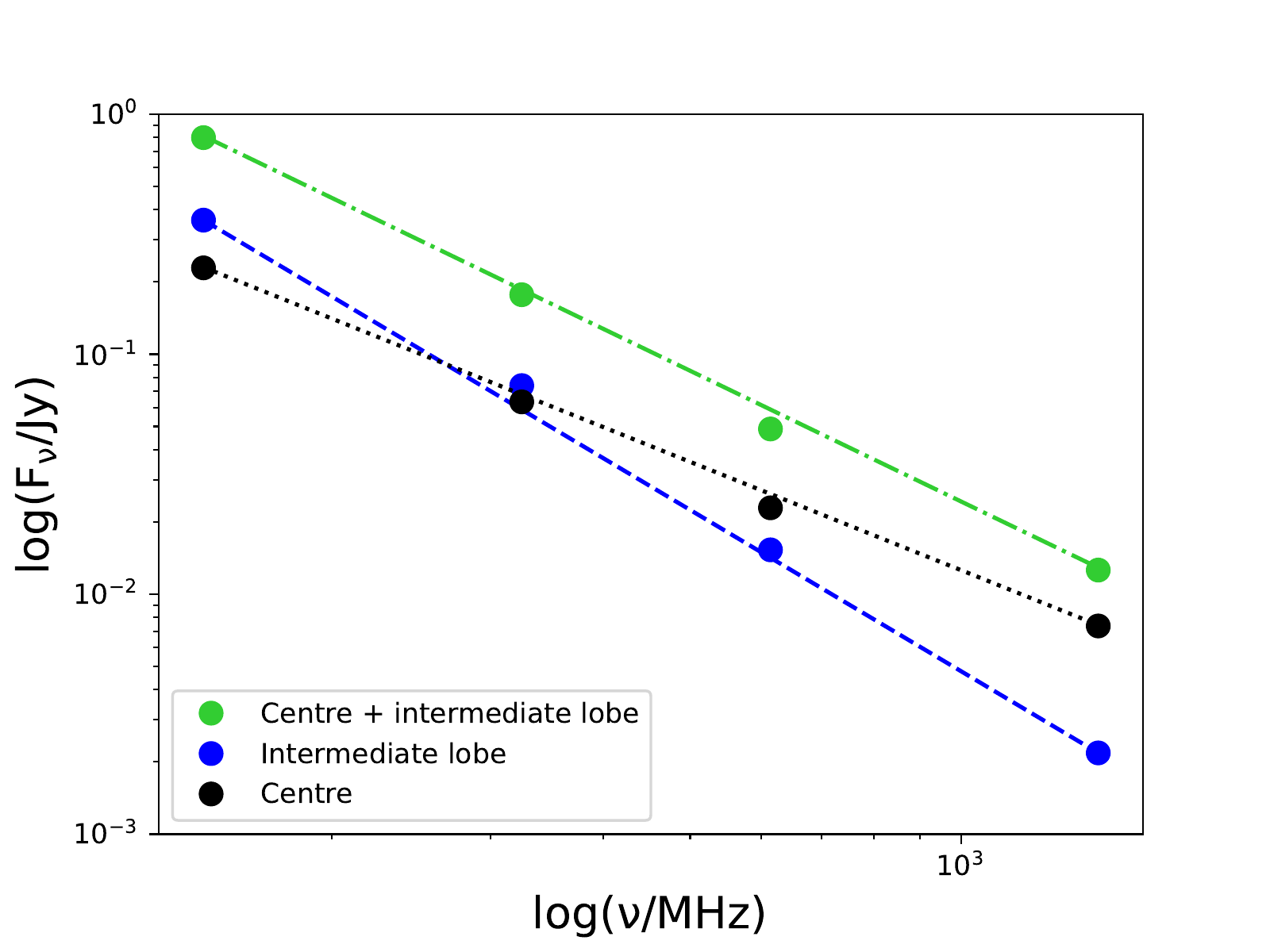}}
    \caption{Integrated flux density as a function of frequency for the central regions shown in Fig. \ref{fig:Regioni}. The centre region is represented with the black dotted line; the intermediate lobe with the blue dashed line, and the total flux of these regions with the green dash-dotted line. In all the cases, the spectra show no evidence of curvature and/or spectral breaks; they are well-described by a power-law with a slope of $\alpha=1.5$ for the centre, $\alpha=2.23$ for the intermediate lobe and $\alpha=1.79$ for centre + intermediate lobe. The size of the points represents the dimension of the flux density error.}
    \label{fig:central_flux}
\end{figure}

\begin{table*}
    \centering
    \renewcommand\arraystretch{1.2}
    \caption{Tribble and CIoff fitting results and correction for adiabatic expansion.}
    \begin{tabular}{cc|ccccc|ccccc}\hline
      \multicolumn{12}{c}{$\alpha_{\rm inj}= 0.75$ \qquad $B_{\rm eq} = 5.9\ \rm{\mu G}$}\\\hline 
     \multicolumn{2}{c}{} &\multicolumn{5}{c}{Tribble model} &\multicolumn{5}{c}{CIoff model}\\
      Region &Age &$\nu_{\rm break}$ &$t$ &$\nu^{\rm corr}_{\rm break}$ &$t^{\rm corr}$ &$\chi^2_{\rm red}$ &$\nu_{\rm break}$ &$t$ &$\nu^{\rm corr}_{\rm break}$ &$t^{\rm corr}$ &$\chi^2_{\rm red}$\\
&&[MHz] &[Myr] &[MHz] &[Myr] & &[MHz] &[Myr] &[MHz] &[Myr] \\\hline
Northern outer lobe &Tot &159 &$170\pm6$ &413 &$106\pm4$  &1.04 &13.6 &$475\pm200$ &0.8 &$2369\pm800$ &4.71\\
&Off &347 &$115\pm7$ &902 &$71\pm4$  &1.95 &466 &$81\pm1$ &28 &$405\pm4$ &$-$\\
&On &$-$&$55\pm15$&$-$&$35\pm8$ &$-$ &$-$ &$394\pm201$ &$-$ &$1964\pm804$ &$-$\\
Southern outer lobe &Tot &166 &$163\pm7$ &481 &$98\pm4$ &1.09 &10.2 &$548\pm200$ &0.6 &$2770\pm800$ &5.68\\
&Off &259 &$133\pm6$ &751 &$78\pm4$ &2.52 &473 &$80\pm1$ &28 &$405\pm4$ &$-$\\
&On &$-$&$30\pm15$&$-$&$20\pm8$ &$-$ &$-$ &$468\pm201$ &$-$ &$2365\pm804$ &$-$\\
Intermediate lobe &Tot &308 &$122\pm7$ &431 &$103\pm6$ &2.60 & $-$ & $-$ & $-$ &$-$ &$-$\\\hline
    \end{tabular}
    \tablefoot{The break frequency ($\nu_{\rm break}$) and the synchrotron age ($t$) derived from the fitting model, the adiabatic-loss corrected break frequency ($\nu_{\rm break}^{\rm corr}$) and synchrotron age ($t^{\rm corr}$) and finally the reduced chi-squared of the fit ($\chi^2_{\rm red})$. Those parameters are derived for both the total age (Tot), the dying phase (Off) and the active phase (On) of the lobes using two different models (Tribble and CIoff), setting $\alpha_{\rm inj}=0.75$ and $B_{\rm eq}=5.9\ \rm{\mu G}$.}
    \label{tab:CIoff}
\end{table*}

\begin{figure}
\resizebox{\hsize}{!}{\includegraphics{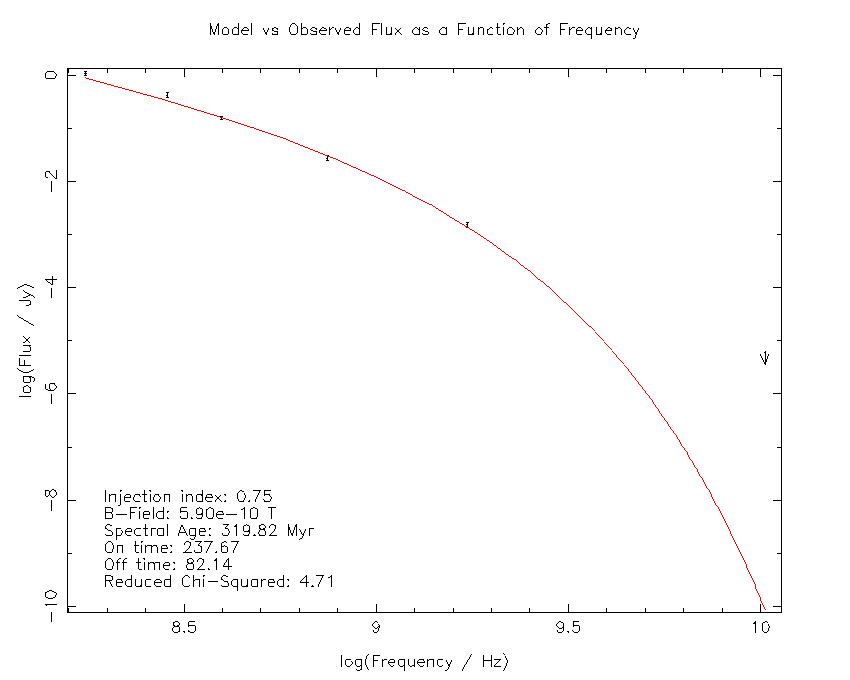}}
    \caption{Results of the CIoff model fit on the northern outer lobe, using $\alpha_{\rm inj}=0.75$.}
    \label{fig:CIoff_fit}
\end{figure}

We selected different sub-regions within the `Centre + intermediate lobe' region depicted in Fig.~\ref{fig:Regioni}, in order to determine whether the intermediate lobe is still being fuelled by the jet. 
The integrated flux densities for these regions are presented in Fig. \ref{fig:central_flux}.
Our results suggest that each sub-region within the inner part of MS0735 is well-described by a single power-law fit, with a slope of $\alpha=1.5$ for the Centre, $\alpha=2.23$ for the intermediate lobe and $\alpha=1.79$ for Centre + intermediate lobe, showing no evidence of a spectral break. This could be explained by ongoing injection of particles, or alternatively through the superposition of various components from different phases. Hence, we can not estimate the spectral ages from the integrated flux densities of these regions.

Conversely, the outer lobes exhibit curved spectra, indicating aged emission, which is consistent with the lack of detection at 8460 MHz (see bottom right panel of Fig. \ref{fig:all_images}).
We fitted the spectra of the outer lobes using the CI and CIoff models, fixing the injection index to the same values as we used for the resolved spectral study ($\alpha_{\rm{inj}}= 0.75$ and 1.5). For both outer lobes we find that the CI model does not provide a good fit to the data, while the CIoff model provides a better fit. The particle supply in the outer lobes is expected to have switched off, as both the radio and X-ray data revealed subsequent phases of jet activity. 
Furthermore, similar to the Tribble model discussed before, the best fit of the CIoff model is obtained with $\alpha_{\rm inj}=1.5$ (see Appendix \ref{sec:appendix} for the results of the fit).

The CIoff model fitting results with $\alpha_{\rm inj}=0.75$ are reported in Table \ref{tab:CIoff}, while the fit of the northern outer lobe flux density is shown in Fig. \ref{fig:CIoff_fit}.
We note that the reduced chi-squared of this fit ($\chi^2_{\rm red}=4.71$) is greater than that obtained from the single-region Tribble model fit ($\chi^2_{\rm red}=2.66$), shown in Fig. \ref{fig:reg_Tribble}. This is true for all the regions in the outer lobes, as is visible from the map in the top right panel of Fig. \ref{fig:Tribble}, where the reduced chi-squared value is always lower than 4. 

As shown in Table \ref{tab:CIoff}, our $\alpha_{\rm inj}=0.75$ fit provides a very low first break frequency ($\nu_{\rm break} \sim 10-15\, {\rm MHz})$  that falls outside the frequency range sampled by our observations. The associated total age of the source ($t\sim0.5\,{\rm Gyr}$) is therefore very old. Even considering the high uncertainties for this value, the discrepancy with the oldest age derived in the resolved spectral study is very large.

\subsection{Correction for adiabatic losses}\label{sec:adiabatic_losses}
We assumed for the spectral ages derived in the previous section that particles only undergo synchrotron and inverse-Compton energy losses.
However, previous X-ray observations have indicated that the cavities filled by the outer lobes were created by low-density bubbles of radio plasma that were inflated by the jet and displaced the surrounding gas by rising buoyantly and expanding adiabatically. Adiabatic losses should therefore be taken into account.

The effect of adiabatic losses is to reduce the particle energy and the magnetic field in radio lobes \citep{Scheuer1968}. The shape of the synchrotron spectrum is preserved, but it is shifted towards lower (or higher) frequencies if there is a single (or continuous) injection of particles \citep{Kardashev1962, Murgia1999}. 
The radiative ages estimated from standard models of spectral ageing without considering the adiabatic expansion exceed (or underestimate) the true age of the source if the synchrotron-emitting electrons are generated in a short-term (or continuous injection) event.

The radiative ages estimated in the previous section can be corrected for by multiplying the break frequency with a corrective factor \emph{F},
\begin{equation}
    \nu_{\rm break}^{corr} = \nu_{\rm break}^{obs} \times \emph{F},
\end{equation}
where for a single injection of particles,
\begin{equation}
    \emph{F} = \biggl(\frac{\Delta^{2n}-1}{2n}\biggr)^2 ,
\end{equation}
while for a continuous injection of particles,
\begin{equation}
    \emph{F} = \frac{1}{4n^2} .
\end{equation}
Here, $\Delta$ is the linear expansion factor for a volume of plasma and $n$ is the exponent of the magnetic field variation $B(t)=B_0\Delta^{-n}=B_0(t/t_0)^{-n}$. 
We set $n=2$, assuming that the magnetic flux is conserved during the expansion.

Following the procedure of \cite{Birzan2008}, we calculated $\Delta$ as the ratio between the final radius of the bubble at the end of the adiabatic expansion ($r_1$) and the initial radius of the
bubble at the start of the adiabatic expansion ($r_0$), which we assumed to be the cluster centre. According to the adiabatic expansion law,
\begin{equation}
    \Delta = \frac{r_1}{r_0} = \biggl(\frac{p_1}{p_0}\biggr)^{-1/3 \gamma},
\end{equation}
where the ratio of specific heats $\gamma$ is assumed to be 4/3, $p_1$ is the gas pressure at the location of the X-ray cavity, and $p_0=5.4\times10^{-10}\ \rm{erg/cm^3}$ is the gas pressure at the centre of the cluster \citep[derived from X-ray observations; see][and Sect. \ref{sec:X}]{Vantyghem2014}.
For the intermediate lobe the detected pressure is less reliable because of the uncertainty about the cavity size and the scatter in the density profile at that distance.

The assumption that the particles are all injected at the central pressure and expand to the current X-ray pressure at each lobe position likely overestimates the adiabatic correction, which is generally modest. This is especially true in our case because the source is at the centre of a cool-core cluster, where the gas pressure is high enough to confine the lobes. This means that the lobes remain luminous for a longer time than remnant lobes in more rarefied environments.

In Table \ref{tab:adiabatic_losses} we list the pressure at the location of the intermediate and outer cavities ($p_1$), the linear expansion factor ($\Delta$), and the corrective factors ($F_{\rm Tribble}$ and $F_{\rm CIoff}$) for the two outer lobes and the intermediate lobe. The corrected break frequencies and ages are reported in Table \ref{tab:CIoff}.

\begin{table}
    \centering
    \renewcommand\arraystretch{1.2}
    \caption{Corrective factors for the adiabatic losses.}
    \begin{tabular}{ccccc} \hline
        Region  &$p_1$ &$\Delta$ &\emph{$F_{\rm Tribble}$}  &\emph{$F_{\rm CIoff}$} \\
        &[$\rm{erg\ cm^{-3}}$] & &&\\\hline
        Northern outer lobe &$7.3\times10^{-11}$ &1.65 &2.6 &0.06\\
        Southern outer lobe &$6.9\times10^{-11}$ &1.67 &2.9 &0.06\\
        Intermediate lobe &$9.5\times10^{-11}$ &1.54 &1.4 & 0.06\\\hline
    \end{tabular}
    \label{tab:adiabatic_losses}
\end{table}

\section{Discussion}\label{sec:discussion}
We have studied the peculiar radio source at the centre of the cool-core cluster MS0735. 
A detailed study of the source in the radio band, together with a re-inspection of the X-ray data, revealed that the source was subjected to at least three different phases of jet activity.
From the radio spectrum we derived estimates of the radiative age of the source under different hypotheses in order to reconstruct its duty cycle and to provide an independent comparison with previously established age estimates from the cavities detected in X-rays. 
We also aimed to estimate the power of the jet to confirm whether the energy released in the ICM during the jet activity is enough to stop the cooling flow, and to verify whether the bubbles are in pressure equilibrium with the surrounding medium.

\subsection{Spectral ages and duty cycle}
A first attempt to derive the spectral age of the source MS0735 was performed by \cite{Birzan2008}. Because data at only two frequencies were available at the time, it was not possible to derive the break frequency from the integrated spectra of the outer lobes. The spectral age was therefore estimated using the relation of \cite{MyersSpangler1985}, assuming $\alpha_{\rm inj}=0.5$. 
The frequency coverage and high resolution of the data presented in this work allowed us to reconstruct the radiative history of the source.

The source MS0735 has experienced at least three different episodes of jet activity throughout its life.
The outer lobes, the best known regions of this source, perfectly match the outer cavities and represent the first phase of jet activity visible today. 
These regions present a very steep and curved spectrum, indicating that particle injection has stopped.
We can therefore give an estimate of their age and reconstruct the duty cycle of the first phase of jet activity.
From the spectral ageing modelling of the northern outer lobe spectrum we derived that the first outburst started around $t_{\rm tot} = 170\pm6$ Myr ago and lasted about $t_{\rm ON} = 55\pm15$ Myr before switch-off ($t_{\rm OFF} = 115\pm7$ Myr).
Similar values are found from the southern outer lobe (see Table \ref{tab:CIoff}).
These numbers were obtained assuming only the presence of synchrotron and inverse-Compton losses and neglecting the adiabatic losses due to the bubble expansion. Therefore the ages reported above could be considered as an upper limit for the true radiative age of the source. 
The age values of the northern outer lobe corrected for adiabatic losses are $t_{\rm tot} = 106\pm3$ Myr, $t_{\rm OFF} = 71\pm3$ Myr, and $t_{\rm ON} = 35\pm6$ Myr (see Sect. \ref{sec:adiabatic_losses}). 
These values were obtained using an upper limit for the lobe expansion and can therefore been considered as a lower limit to the actual lobe age.
In summary, the total age of the outer lobes lies in the range $106\,{\rm Myr} \lesssim t_{\rm tot} \lesssim 170\,{\rm Myr,}$ and the first phase of jet activity lasted around $35\,{\rm Myr} \lesssim t_{\rm ON} \lesssim 55\,{\rm Myr}$.

The second phase of jet activity is associated with the intermediate lobe, which also presents an overall steep spectrum, and fills the new discovered intermediate cavity.
The age gradient in this region cannot be well constrained because the available resolution is too low. Therefore we only provide an estimate of the start of the second phase of jet activity, which lies in the range $103\,{\rm Myr} \lesssim t_{\rm tot} \lesssim 122\,{\rm Myr}$, depending on the contribution of the adiabatic expansion.
We note that these values are comparable with the age of the outer lobes, indicating that jet activity has immediately been reactivated, with an inactivity time of about a few and 10 Myr.

The third phase of the jet activity is represented by the inner lobes, for which we are not able to give an estimate of the age, however, because they are not resolved.
The central regions of the source are poorly described by the Tribble model, as indicated by the high chi-squared values we obtained (lower right panel of Fig. \ref{fig:Tribble}). They result from the likely superposition of particles that were emitted at different times.
However, the measurement at high resolution of a central spectral index of 0.75 suggests active jets on scales of some kiloparsec. Furthermore, the detection of inner cavities on the same scale ensures the reactivation of jet activity.
The age of the inner cavities could then be used to give a constraint on the duration of the second phase.
Based on the age of the southern inner cavity \citep[25$-$78 Myr, depending on the scenario,][]{Vantyghem2014} and that of the intermediate lobe, the jet cannot be active for more than $t_{\rm ON} \le 44 - 78$ Myr.

Overall, our new analysis suggests that the source is going through a very rapid cycle where the central AGN is active most of the time, interrupted only by brief quiescent phases.
These several close episodes of jet activity ensure a continuous source of heating to the central gas, preventing it from cooling.

\begin{table*}
    \centering
    \renewcommand\arraystretch{1.2}
    \caption{Radio lobe versus X-ray cavity age estimates.}
    \begin{tabular}{ccccc} \hline
        Region &$t_{\rm rad}$ &$t_{\rm buoyancy}$ &$t_{\rm Cs}$ &$t_{\rm refill}$ \\
       &[Myr]&[Myr] &[Myr] &[Myr] \\\hline
        Northern outer lobe/ Northern outer cavity &106$-$170  &91 &120 &240\\
        Southern outer lobe/ Southern outer cavity &98$-$163  &110 &140 &250\\
        Intermediate lobe/ Intermediate cavity &103$-$122 &62-64 &53-69 &103-120 \\\hline
    \end{tabular}
    \label{tab:Comparison}
    \tablefoot{For the age of the radio lobe ($t_{\rm rad}$) derived from our analysis we report a range of values. The highest value is derived with a simple radiative Tribble model fit, while the lowest is corrected for adiabatic losses. The three X-ray cavity age estimates are based on the buoyancy timescale $(t_{\rm{buoyancy}})$, the sound crossing timescale $(t_{\rm{Cs}}),$ and the refill timescale $(t_{\rm{refill}})$. The southern and northern outer cavity timescales are derived by \cite{Vantyghem2014}, while we estimate the timescales for the intermediate cavity reporting a range of values for the different dimensions of the cavity considered.}
\end{table*}

We recall that our discussion of the source age and duty cycle is based on the results of the resolved spectral study, which we consider more reliable.
The analysis of the integrated spectrum has provided less certain results. The total times derived from the fit of the CIoff model are three times higher than those derived from the resolved study and are therefore unrealistic. 
Furthermore, the chi-squared values of the CIoff fit for the integrated spectra of the outer lobes are higher than those obtained from the Tribble model fit for single regions. \cite{Harwood2017a} reported that a discrepancy between the spectral ages derived from integrated and resolved analysis is expected. An integrated study is not able to provide good constraints with respect to the model parameters.
Furthermore, the goodness-of-fit is highly dependent on frequency coverage if the spectrum is not well constrained around the spectral break, as is our case for the first break frequency. 

When compared with other radio galaxies, the timescales we observe for MS0735 are in accordance with the typical activity time, between dozens and some hundred million years, and the inactivity time is between a few and some dozen million years \citep[e.g.][]{Konar2013,Orru2015,Shulevski2017,Brienza2018,Brienza2020,Maccagni2020}.
Unfortunately, not many similar studies on radio galaxies at the centre of clusters are available, so we cannot make a direct comparison. The only clusters that were studied are 3C438 and 3C28 \citep[][]{Harwood2015}, which present a single phase of jet activity and are younger (oldest age in the lobes of $\sim4$ and $\sim17$ Myr, respectively), and 3C388 \citep[][]{Brienza2020}, which instead is a restarted source with a total age of 82 Myr and a longer activity time than the quiescent period, as seen in our source. 
However, the sample is not large enough to draw general conclusions.

\subsection{Comparison with X-ray estimates}
An independent estimate of the source age was made by \cite{Vantyghem2014} from the X-ray data, providing different timescales for the formation of the outer cavities. 
We can then compare the total radiative age we derived for the outer lobes with the age derived from X-rays for the corresponding cavities. To our knowledge, no studies of sources in clusters are available in which this type of comparison has been made. It was only performed for a source in a galaxy group \citep[][]{Kolokythas2020}. It is therefore important to understand whether the two estimates are consistent, as Kolokythas and collaborators have found.

Three different timescales were used to derive the age of the cavities: the buoyancy, sound crossing, and refill timescale. In general, the ages calculated using the speed of sound are the shortest, those based on the refilling timescale are the longest, and the buoyancy timescale, which is based on the terminal velocity of the bubbles, lies in between. However, in the case of the outer cavities of MS0735, the terminal velocity is supersonic, so that the buoyancy timescale is the shortest. Neglecting the expansion history of the bubble in the buoyancy timescale therefore underestimates the true cavity age.
All timescales are reported in Table \ref{tab:Comparison}.

When we compared the age estimates of the outer lobes or cavities,
we realised that the buoyancy (91 Myr) and sound crossing timescales (120 Myr) are comparable with the derived range of radiative ages (106$-$170 Myr), while the refilling timescale (240 Myr) is much longer. We note that it appears from the X-ray analysis that the refilling timescale is overestimated because it is longer than the age of the surrounding shock front $(t_{\rm shock}=110\,{\rm Myr})$, which should be comparable to the true cavity age \citep[][]{Vantyghem2014}.
We therefore suggest that the refilling scenario is the least reliable.

A comparison can also be made for the age of the intermediate lobe or cavity. We followed the procedure used by \cite{Vantyghem2014} to compute the three different timescales for this new cavity, reported in Table \ref{tab:Comparison}.
In contrast to the outer lobes, the range of radiative ages that we found for this lobe (103$-$122 Myr) is comparable to the estimates provided by the refilling scenario.

For the inner cavities, on the other hand, we cannot make comparable statements as we were unable able to estimate the age of the radio emission on the same scale.
Overall, we note that the two ways of estimating the age of a source give consistent results. 
When the radiative age is better constrained, it allows reducing the number of methods for calculating the age of the cavity.
However, our results are not sufficient to determine which is the most realistic.

\subsection{Energetics}
In this section we estimate the energetics of the source, starting from the radio properties of the outer lobes.
Deriving the total power of the source is important to understand whether the gas heating by AGN is enough to quench the cooling flow in the cluster centre, as is observed from the X-rays.
The X-ray analysis of this source has revealed that the energy released by the AGN outburst ($\sim8\times10^{61}$ erg) is enough to quench a cooling flow for several billion years \citep{McNamara2005,Gitti2007,Vantyghem2014}.
When we take the mean of the buoyancy timescales of the outer cavities ($t\sim100$ Myr) as an estimate of the cavity age, the mechanical power required to inflate the bubbles is $P = E/t = 1.4\times10^{46}\ \rm{erg\ s^{-1}}$.
The same estimate can be performed based on the radio data presented in this work as follows.

It is difficult to estimate the total power of the source because different components contribute to it.
First of all, we can compute the total radio luminosity for the outer lobes by integrating the flux between the rest-frame frequencies of 10 and 10000 MHz as
\begin{equation}
    L_{\rm rad} = \frac{4\pi D_{\rm L}^2S_{\nu_0}}{(1+z)^{(1-\alpha)}}\int_{\nu_1}^{\nu_2} (\nu/\nu_0)^{-\alpha} d\nu,
\end{equation}
where we used the flux density measured at 144 MHz as the flux reference $(S_{\nu_0})$, and set the spectral index to $\alpha=2.9$. This is the value derived in the outer lobes, using our data at 144 MHz and 610 MHz (because this spectral index is probably measured above the break frequency, integrating to low frequency will overestimate the luminosity). We obtain a total radio luminosity of $L_{\rm rad}=7\times10^{43}\ \rm{erg\ s^{-1}}$. 

The energy dissipated into synchrotron radiation is only a minor fraction of the jet power. The bulk of the jet power is converted into energizing the lobes, which contain both relativistic particles and a magnetic field. The total energy is hence given by the sum of these contributions $E_{\rm tot} = (1+k)E_{\rm e} + E_{\rm B}$, where k is the ratio between the energy of protons and electrons, which we have set equal to one. Both the particle and the magnetic field energy depend on the magnetic field strength, which is difficult to measure, so that an estimate can be derived by invoking minimum-energy arguments. The condition of minimum energy approximately corresponds to energy equipartition between particles and magnetic field. For the equipartition magnetic field strength derived in Sect. \ref{sec:radiative_age}, $B_{\rm eq}\sim 5.9\ \upmu\rm{G}$, the minimum energy contained in the lobes is $E_{\rm min} \sim 9\times10^{59}\ \rm{erg}$. Furthermore, we can use our estimate of the lobe age to determine the average jet power over the lifetime of MS0735.

Using the total age of northern outer lobe ($t_{\rm tot}=62-170$ Myr), we derive a range for the power $P_{\rm jet}\sim 2-5\times10^{44}\ \rm{erg\ s^{-1}}$, which is two orders of magnitude lower than the power derived from the X-rays and comparable with the X-ray bolometric luminosity within the cooling radius ($2.6\times10^{44}\ \rm{erg\ s^{-1}}$), balancing radiative losses. However, we must consider that in our calculation we used the minimum energy inside the lobes, which means that the power that is injected might be larger. Furthermore, neither estimate accounts for the power emitted to drive shocks into the external medium, as observed from X-ray data.  

From the minimum energy we can also estimate the minimum pressure inside the lobes, the sum of the contribution of the magnetic field and of the plasma (for a relativistic plasma $p=U/3$, where U is the energy density). This yields $p\sim 2\times10^{-12} \, \rm{erg\ cm^{-3}}$, which is again two orders of magnitude lower than the gas pressure of the external medium derived from the X-rays ($\sim2\times10^{-10}\ \rm{erg\ cm^{-3}}$). This is clearly unphysical because in this case, the bubbles would collapse under the pressure of the external medium.

In our estimates we considered that the particle energy is equally divided between electrons and protons (k=1) and that the lobes are completely filled with radio plasma (filling factor $\phi=1$). 
To reach pressure equilibrium, keeping the equipartition principle valid, we can decrease the filling factor and increase the contribution of the protons. In this case, we would assume $k\ge2000$ and $\phi\le30\%$, generating much higher magnetic field strengths ($B_{\rm eq}\sim 60\ \rm{\mu G}$) and therefore unrealistic short plasma ages ($\sim 3$ Myr).
Furthermore, the strongest evidence that this is not the case is given by the fact that FRII radio galaxies have light jets \citep{Hardcastle2002,Croston2004,Croston2005,Croston2018}. 
This means that either equipartition does not apply in the outer lobes of MS0735 or that there is additional pressure support (e.g. from entraining of hot thermal gas).

In the case of FRII sources, where it was possible to directly measure the magnetic field strength from observations of inverse-Compton emission, it has been shown that the typical magnetic field strength is a factor 2-3 below the equipartition values \cite[e.g.][]{Croston2005,Kataoka2005,Migliori2007,Ineson2017,Turner2018}, suggesting that the lobes contain electron energy densities higher than what is implied by the minimum energy condition. 
Accordingly, the total energy would be higher than the minimum energy by a factor of about 2 on average. 
However, the departure from equipartition is not sufficient to provide the solution to the underpressured lobes of MS0735.

In conclusion, additional pressure support is required to keep bubbles in equilibrium. This could be provided by hot thermal gas entrained by the jet, as has been proposed to compensate for the large pressure disparity observed in a few other sources, for example the intermediate FRI/FRII source Hydra A \citep[][]{Croston2014} and the FRII source 3C438 \citep[][]{Harwood2015}.
To entrain material, a jet must move slowly, but the material of FRII sources is assumed to remain relativistic up to great distances. However, MS0735 does not show the typical morphology of an FRII, as already noted. A slow-moving jet might therefore provide a solution to both the pressure gap and its atypical morphology.

\section{Conclusions}\label{sec:conclusions}
We have presented new LOFAR observations of the source MS 0735.6+7421. These data, combined with archival GMRT and VLA data at higher frequencies, allowed us to perform a resolved spectral study of this source, expanding previous work both in frequency range and in resolution.
The main results are summarised below.
\begin{itemize}
\item Our new LOFAR data show two giant outer radio lobes, much wider than at higher frequencies, with a hotspot at the extremity. These lobes fill the X-ray outer cavities perfectly. There is also evidence of an intermediate lobe south of the core, confirmed by our discovery of a corresponding cavity after a re-inspection of the X-ray data.
This means that the source experienced a further phase of jet activity, in addition to the two activities that were known before from a previous analysis of X-ray data.
\item Previous work has pointed out the steep spectrum for this source. Our work confirms this result. The spectrum is steep down to 144 MHz both in the outer lobes ($\alpha_{144}^{610}\sim2.9$) and in the more central regions ($\alpha_{144}^{610}\sim2.1$). Additionally, the hotspots in the outer lobes show a steep spectral index ($\alpha_{144}^{610}\sim2.4$), indicating that the particle acceleration mechanisms in these regions has ceased to operate. However, the spectral index flattens at low frequencies to a value of $\alpha_{144}^{235}\sim0.75$.
\item Our high-resolution images of the central emission at 1420 MHz and 8460 MHz show a spectral index of $\alpha_{1420}^{8460}=0.75\pm0.08$. 
This suggests that within the central component, the actual AGN core and the inner active jets are blended.
\item We used single-injection models to estimate the radiative age of the radio lobes and found a total age for the outer lobes in the range 106-170 Myr and for the intermediate lobe in the range 103-122 Myr, depending on the effect of adiabatic losses. We compared radiative ages with X-ray estimates from cavity timescales. The age range of the outer lobes we found agrees well with buoyancy and sound crossing timescales derived from X-rays \citep{Vantyghem2014}. 
\item We reconstructed the duty cycle of the source, finding that the first phase of jet activity lasted for about $t_{\rm ON}= 35-55$ Myr. We also set an upper limit on the duration of the second phase ($t_{\rm ON} \le 44-78$ Myr). The two episodes are separated by a brief quiescent phase of a few million years. The source therefore has a duty cycle close to unity, and the AGN was active for most of the time.
\item By comparing the pressure estimates inside the cavities, derived from the minimum total energy and from the thermal gas, we found that additional pressure support, for instance from entraining of hot thermal gas, is necessary to maintain the bubbles.
\end{itemize}

We have demonstrated the importance of performing a resolved spectral study to reconstruct the duty cycle of a radio source. This is necessary to understand the AGN feedback. 
With the high resolution and sensitivity reached with the new generation of radio telescopes (LOFAR, the Karl G. Jansky VLA, the uGMRT, and in the near future with the advent of the Square Kilometre Array; SKA), resolved spectral studies will be enabled for an increasing number of sources, allowing us to perform the first statistical studies of these objects. It will also shed new light on the history of very well-known sources such as MS0735.

\section*{Acknowledgements}
We thank the Referee for helpful comments.
NB, MB, AB, EB, CJR acknowledge support from the ERC through the grant ERC-Stg DRANOEL n. 714245. 
ACE acknowledges support from STFC grant ST/P00541/1. 
LOFAR, the Low Frequency Array designed and constructed by ASTRON (Netherlands Institute for Radio Astronomy), has facilities in several countries, that are owned by various parties (each with their own funding sources), and that are collectively operated by the International LOFAR Telescope (ILT) foundation under a joint scientific policy.
The scientific results reported in this article are based in part on data obtained from the VLA Data Archive, the GMRT Data Archive and the \emph{Chandra} Data Archive. 
This research made use of different packages for pyhton: APLpy \citep{Robitaille2012}, Astropy \citep{Astropy2013}, NumPy \citep{vanderwalt2011} and SciPy \citep{Scipy2020}.

\bibliographystyle{aa}
\interlinepenalty=10000
\bibliography{MS0735}

\begin{thebibliography}{103}
\expandafter\ifx\csname natexlab\endcsname\relax\def\natexlab#1{#1}\fi

\bibitem[{{Astropy Collaboration} {et~al.}(2013){Astropy Collaboration},
  {Robitaille}, {Tollerud}, {Greenfield}, {Droettboom}, {Bray}, {Aldcroft},
  {Davis}, {Ginsburg}, {Price-Whelan}, {Kerzendorf}, {Conley}, {Crighton},
  {Barbary}, {Muna}, {Ferguson}, {Grollier}, {Parikh}, {Nair}, {Unther},
  {Deil}, {Woillez}, {Conseil}, {Kramer}, {Turner}, {Singer}, {Fox}, {Weaver},
  {Zabalza}, {Edwards}, {Azalee Bostroem}, {Burke}, {Casey}, {Crawford},
  {Dencheva}, {Ely}, {Jenness}, {Labrie}, {Lim}, {Pierfederici}, {Pontzen},
  {Ptak}, {Refsdal}, {Servillat}, \& {Streicher}}]{Astropy2013}
{Astropy Collaboration}, {Robitaille}, T.~P., {Tollerud}, E.~J., {et~al.} 2013,
  \aap, 558, A33

\bibitem[{{B{\^\i}rzan} {et~al.}(2008){B{\^\i}rzan}, {McNamara}, {Nulsen},
  {Carilli}, \& {Wise}}]{Birzan2008}
{B{\^\i}rzan}, L., {McNamara}, B.~R., {Nulsen}, P.~E.~J., {Carilli}, C.~L., \&
  {Wise}, M.~W. 2008, \apj, 686, 859

\bibitem[{{B{\^\i}rzan} {et~al.}(2004){B{\^\i}rzan}, {Rafferty}, {McNamara},
  {Wise}, \& {Nulsen}}]{Birzan2004}
{B{\^\i}rzan}, L., {Rafferty}, D.~A., {McNamara}, B.~R., {Wise}, M.~W., \&
  {Nulsen}, P.~E.~J. 2004, \apj, 607, 800

\bibitem[{{Blandford} \& {K{\"o}nigl}(1979)}]{Blandford1979}
{Blandford}, R.~D. \& {K{\"o}nigl}, A. 1979, \apj, 232, 34

\bibitem[{{Brienza} {et~al.}(2020){Brienza}, {Morganti}, {Harwood}, {Duchet},
  {Rajpurohit}, {Shulevski}, {Hardcastle}, {Mahatma}, {Godfrey}, {Prandoni},
  {Shimwell}, \& {Intema}}]{Brienza2020}
{Brienza}, M., {Morganti}, R., {Harwood}, J., {et~al.} 2020, \aap, 638, A29

\bibitem[{{Brienza} {et~al.}(2018){Brienza}, {Morganti}, {Murgia}, {Vilchez},
  {Adebahr}, {Carretti}, {Concu}, {Govoni}, {Harwood}, {Intema}, {Loi},
  {Melis}, {Paladino}, {Poppi}, {Shulevski}, {Vacca}, \&
  {Valente}}]{Brienza2018}
{Brienza}, M., {Morganti}, R., {Murgia}, M., {et~al.} 2018, \aap, 618, A45

\bibitem[{{Carilli} {et~al.}(1991){Carilli}, {Perley}, {Dreher}, \&
  {Leahy}}]{Carilli1991}
{Carilli}, C.~L., {Perley}, R.~A., {Dreher}, J.~W., \& {Leahy}, J.~P. 1991,
  \apj, 383, 554

\bibitem[{{Chandra} {et~al.}(2004){Chandra}, {Ray}, \&
  {Bhatnagar}}]{Chandra2004}
{Chandra}, P., {Ray}, A., \& {Bhatnagar}, S. 2004, \apj, 612, 974

\bibitem[{{Cohen} {et~al.}(2005){Cohen}, {Clarke}, {Feretti}, \&
  {Kassim}}]{Cohen2005}
{Cohen}, A.~S., {Clarke}, T.~E., {Feretti}, L., \& {Kassim}, N.~E. 2005, \apjl,
  620, L5

\bibitem[{{Croston} {et~al.}(2004){Croston}, {Birkinshaw}, {Hardcastle}, \&
  {Worrall}}]{Croston2004}
{Croston}, J.~H., {Birkinshaw}, M., {Hardcastle}, M.~J., \& {Worrall}, D.~M.
  2004, \mnras, 353, 879

\bibitem[{{Croston} \& {Hardcastle}(2014)}]{Croston2014}
{Croston}, J.~H. \& {Hardcastle}, M.~J. 2014, \mnras, 438, 3310

\bibitem[{{Croston} {et~al.}(2005){Croston}, {Hardcastle}, {Harris}, {Belsole},
  {Birkinshaw}, \& {Worrall}}]{Croston2005}
{Croston}, J.~H., {Hardcastle}, M.~J., {Harris}, D.~E., {et~al.} 2005, \apj,
  626, 733

\bibitem[{{Croston} {et~al.}(2018){Croston}, {Ineson}, \&
  {Hardcastle}}]{Croston2018}
{Croston}, J.~H., {Ineson}, J., \& {Hardcastle}, M.~J. 2018, \mnras, 476, 1614

\bibitem[{{De Young}(1984)}]{deYoung1984}
{De Young}, D.~S. 1984, \physrep, 111, 373

\bibitem[{{de Young}(2002)}]{deYoung2002}
{de Young}, D.~S. 2002, {The physics of extragalactic radio sources}

\bibitem[{{Donahue} \& {Stocke}(1995)}]{DonahueStocke1995}
{Donahue}, M. \& {Stocke}, J.~T. 1995, \apj, 449, 554

\bibitem[{{Donahue} {et~al.}(1992){Donahue}, {Stocke}, \&
  {Gioia}}]{Donahue1992}
{Donahue}, M., {Stocke}, J.~T., \& {Gioia}, I.~M. 1992, \apj, 385, 49

\bibitem[{{Doria} {et~al.}(2012){Doria}, {Gitti}, {Ettori}, {Brighenti},
  {Nulsen}, \& {McNamara}}]{Doria2012}
{Doria}, A., {Gitti}, M., {Ettori}, S., {et~al.} 2012, \apj, 753, 47

\bibitem[{{Dunn} {et~al.}(2006){Dunn}, {Fabian}, \& {Celotti}}]{Dunn2006}
{Dunn}, R.~J.~H., {Fabian}, A.~C., \& {Celotti}, A. 2006, \mnras, 372, 1741

\bibitem[{{Fabian}(1994)}]{Fabian1994}
{Fabian}, A.~C. 1994, \araa, 32, 277

\bibitem[{{Fabian} {et~al.}(2000){Fabian}, {Sanders}, {Ettori}, {Taylor},
  {Allen}, {Crawford}, {Iwasawa}, {Johnstone}, \& {Ogle}}]{Fabian2000}
{Fabian}, A.~C., {Sanders}, J.~S., {Ettori}, S., {et~al.} 2000, \mnras, 318,
  L65

\bibitem[{{Falcke} \& {Biermann}(1995)}]{Falcke1995}
{Falcke}, H. \& {Biermann}, P.~L. 1995, \aap, 293, 665

\bibitem[{{Fanaroff} \& {Riley}(1974)}]{FR1974}
{Fanaroff}, B.~L. \& {Riley}, J.~M. 1974, \mnras, 167, 31P

\bibitem[{{Forman} \& {Jones}(1982)}]{Forman1982}
{Forman}, W. \& {Jones}, C. 1982, \araa, 20, 547

\bibitem[{{Giacintucci} {et~al.}(2020){Giacintucci}, {Markevitch},
  {Johnston-Hollitt}, {Wik}, {Wang}, \& {Clarke}}]{Giacintucci2020}
{Giacintucci}, S., {Markevitch}, M., {Johnston-Hollitt}, M., {et~al.} 2020,
  \apj, 891, 1

\bibitem[{{Gitti} {et~al.}(2006){Gitti}, {Feretti}, \& {Schindler}}]{Gitti2006}
{Gitti}, M., {Feretti}, L., \& {Schindler}, S. 2006, \aap, 448, 853

\bibitem[{{Gitti} {et~al.}(2007){Gitti}, {McNamara}, {Nulsen}, \&
  {Wise}}]{Gitti2007}
{Gitti}, M., {McNamara}, B.~R., {Nulsen}, P.~E.~J., \& {Wise}, M.~W. 2007,
  \apj, 660, 1118

\bibitem[{{Gitti} {et~al.}(2011){Gitti}, {Nulsen}, {David}, {McNamara}, \&
  {Wise}}]{Gitti2011}
{Gitti}, M., {Nulsen}, P. E.~J., {David}, L.~P., {McNamara}, B.~R., \& {Wise},
  M.~W. 2011, \apj, 732, 13

\bibitem[{{Gitti} {et~al.}(2010){Gitti}, {O'Sullivan}, {Giacintucci}, {David},
  {Vrtilek}, {Raychaudhury}, \& {Nulsen}}]{Gitti2010}
{Gitti}, M., {O'Sullivan}, E., {Giacintucci}, S., {et~al.} 2010, \apj, 714, 758

\bibitem[{{Hardcastle}(2013)}]{Hardcastle2013}
{Hardcastle}, M.~J. 2013, \mnras, 433, 3364

\bibitem[{{Hardcastle} {et~al.}(1998){Hardcastle}, {Alexander}, {Pooley}, \&
  {Riley}}]{Hardcastle1998}
{Hardcastle}, M.~J., {Alexander}, P., {Pooley}, G.~G., \& {Riley}, J.~M. 1998,
  \mnras, 296, 445

\bibitem[{{Hardcastle} {et~al.}(2002){Hardcastle}, {Birkinshaw}, {Cameron},
  {Harris}, {Looney}, \& {Worrall}}]{Hardcastle2002}
{Hardcastle}, M.~J., {Birkinshaw}, M., {Cameron}, R.~A., {et~al.} 2002, \apj,
  581, 948

\bibitem[{{Hardcastle} \& {Croston}(2020)}]{HardcastleCroston2020}
{Hardcastle}, M.~J. \& {Croston}, J.~H. 2020, arXiv e-prints, arXiv:2003.06137

\bibitem[{{Hardcastle} {et~al.}(2020){Hardcastle}, {Shimwell}, {Tasse}, {Best},
  {Drabent}, {Jarvis}, {Prandoni}, {Rottgering}, {Sabater}, \&
  {Schwarz}}]{Hardcastle2020}
{Hardcastle}, M.~J., {Shimwell}, T.~W., {Tasse}, C., {et~al.} 2020, arXiv
  e-prints, arXiv:2011.08294

\bibitem[{{Harwood}(2017a)}]{Harwood2017a}
{Harwood}, J.~J. 2017a, \mnras, 466, 2888

\bibitem[{{Harwood} {et~al.}(2015){Harwood}, {Hardcastle}, \&
  {Croston}}]{Harwood2015}
{Harwood}, J.~J., {Hardcastle}, M.~J., \& {Croston}, J.~H. 2015, \mnras, 454,
  3403

\bibitem[{{Harwood} {et~al.}(2013){Harwood}, {Hardcastle}, {Croston}, \&
  {Goodger}}]{Harwood2013}
{Harwood}, J.~J., {Hardcastle}, M.~J., {Croston}, J.~H., \& {Goodger}, J.~L.
  2013, \mnras, 435, 3353

\bibitem[{{Harwood} {et~al.}(2017b){Harwood}, {Hardcastle}, {Morganti},
  {Croston}, {Br{\"u}ggen}, {Brunetti}, {R{\"o}ttgering}, {Shulevski}, \&
  {White}}]{Harwood2017b}
{Harwood}, J.~J., {Hardcastle}, M.~J., {Morganti}, R., {et~al.} 2017b, \mnras,
  469, 639

\bibitem[{{Heald} {et~al.}(2015){Heald}, {Pizzo}, {Orr{\'u}}, {Breton},
  {Carbone}, {Ferrari}, {Hardcastle}, {Jurusik}, {Macario}, {Mulcahy},
  {Rafferty}, {Asgekar}, {Brentjens}, {Fallows}, {Frieswijk}, {Toribio},
  {Adebahr}, {Arts}, {Bell}, {Bonafede}, {Bray}, {Broderick}, {Cantwell},
  {Carroll}, {Cendes}, {Clarke}, {Croston}, {Daiboo}, {de Gasperin}, {Gregson},
  {Harwood}, {Hassall}, {Heesen}, {Horneffer}, {van der Horst}, {Iacobelli},
  {Jeli{\'c}}, {Jones}, {Kant}, {Kokotanekov}, {Martin}, {McKean}, {Morabito},
  {Nikiel-Wroczy{\'n}ski}, {Offringa}, {Pandey}, {Pandey-Pommier}, {Pietka},
  {Pratley}, {Riseley}, {Rowlinson}, {Sabater}, {Scaife}, {Scheers},
  {Sendlinger}, {Shulevski}, {Sipior}, {Sobey}, {Stewart}, {Stroe}, {Swinbank},
  {Tasse}, {Tr{\"u}stedt}, {Varenius}, {van Velzen}, {Vilchez}, {van Weeren},
  {Wijnholds}, {Williams}, {de Bruyn}, {Nijboer}, {Wise}, {Alexov}, {Anderson},
  {Avruch}, {Beck}, {Bell}, {van Bemmel}, {Bentum}, {Bernardi}, {Best},
  {Breitling}, {Brouw}, {Br{\"u}ggen}, {Butcher}, {Ciardi}, {Conway}, {de
  Geus}, {de Jong}, {de Vos}, {Deller}, {Dettmar}, {Duscha}, {Eisl{\"o}ffel},
  {Engels}, {Falcke}, {Fender}, {Garrett}, {Grie{\ss}meier}, {Gunst},
  {Hamaker}, {Hessels}, {Hoeft}, {H{\"o}rand el}, {Holties}, {Intema},
  {Jackson}, {J{\"u}tte}, {Karastergiou}, {Klijn}, {Kondratiev}, {Koopmans},
  {Kuniyoshi}, {Kuper}, {Law}, {van Leeuwen}, {Loose}, {Maat}, {Markoff},
  {McFadden}, {McKay-Bukowski}, {Mevius}, {Miller-Jones}, {Morganti}, {Munk},
  {Nelles}, {Noordam}, {Norden}, {Paas}, {Polatidis}, {Reich}, {Renting},
  {R{\"o}ttgering}, {Schoenmakers}, {Schwarz}, {Sluman}, {Smirnov}, {Stappers},
  {Steinmetz}, {Tagger}, {Tang}, {ter Veen}, {Thoudam}, {Vermeulen}, {Vocks},
  {Vogt}, {Wijers}, {Wucknitz}, {Yatawatta}, \& {Zarka}}]{Heald15}
{Heald}, G.~H., {Pizzo}, R.~F., {Orr{\'u}}, E., {et~al.} 2015, \aap, 582, A123

\bibitem[{{Hogan} {et~al.}(2015){Hogan}, {Edge}, {Hlavacek-Larrondo},
  {Grainge}, {Hamer}, {Mahony}, {Russell}, {Fabian}, {McNamara}, \&
  {Wilman}}]{Hogan2015}
{Hogan}, M.~T., {Edge}, A.~C., {Hlavacek-Larrondo}, J., {et~al.} 2015, \mnras,
  453, 1201

\bibitem[{{Hogan} {et~al.}(2017){Hogan}, {McNamara}, {Pulido}, {Nulsen},
  {Vantyghem}, {Russell}, {Edge}, {Babyk}, {Main}, \& {McDonald}}]{Hogan2017}
{Hogan}, M.~T., {McNamara}, B.~R., {Pulido}, F.~A., {et~al.} 2017, \apj, 851,
  66

\bibitem[{{Ineson} {et~al.}(2017){Ineson}, {Croston}, {Hardcastle}, \&
  {Mingo}}]{Ineson2017}
{Ineson}, J., {Croston}, J.~H., {Hardcastle}, M.~J., \& {Mingo}, B. 2017,
  \mnras, 467, 1586

\bibitem[{{Intema}(2014)}]{Intema2014}
{Intema}, H.~T. 2014, in Astronomical Society of India Conference Series,
  Vol.~13, Astronomical Society of India Conference Series, 469

\bibitem[{{Intema} {et~al.}(2017){Intema}, {Jagannathan}, {Mooley}, \&
  {Frail}}]{Intema2017}
{Intema}, H.~T., {Jagannathan}, P., {Mooley}, K.~P., \& {Frail}, D.~A. 2017,
  \aap, 598, A78

\bibitem[{{Jaffe} \& {Perola}(1973)}]{JaffePerola1973}
{Jaffe}, W.~J. \& {Perola}, G.~C. 1973, \aap, 26, 423

\bibitem[{{Kaiser} \& {Best}(2007)}]{KaiserBest2007}
{Kaiser}, C.~R. \& {Best}, P.~N. 2007, \mnras, 381, 1548

\bibitem[{{Kapi{\'n}ska} {et~al.}(2017){Kapi{\'n}ska}, {Terentev}, {Wong},
  {Shabala}, {Andernach}, {Rudnick}, {Storer}, {Banfield}, {Willett}, {de
  Gasperin}, {Lintott}, {L{\'o}pez-S{\'a}nchez}, {Middelberg}, {Norris},
  {Schawinski}, {Seymour}, \& {Simmons}}]{Kapinska2017}
{Kapi{\'n}ska}, A.~D., {Terentev}, I., {Wong}, O.~I., {et~al.} 2017, \aj, 154,
  253

\bibitem[{{Kardashev}(1962)}]{Kardashev1962}
{Kardashev}, N.~S. 1962, \sovast, 6, 317

\bibitem[{{Kataoka} \& {Stawarz}(2005)}]{Kataoka2005}
{Kataoka}, J. \& {Stawarz}, {\L}. 2005, \apj, 622, 797

\bibitem[{{Kellermann} \& {Owen}(1988)}]{Kellermann1988}
{Kellermann}, K.~I. \& {Owen}, F.~N. 1988, {Radio galaxies and quasars.}, ed.
  K.~I. {Kellermann} \& G.~L. {Verschuur}, 563--602

\bibitem[{{Kokotanekov} {et~al.}(2017){Kokotanekov}, {Wise}, {Heald}, {McKean},
  {B{\^\i}rzan}, {Rafferty}, {Godfrey}, {de Vries}, {Intema}, {Broderick},
  {Hardcastle}, {Bonafede}, {Clarke}, {van Weeren}, {R{\"o}ttgering}, {Pizzo},
  {Iacobelli}, {Orr{\'u}}, {Shulevski}, {Riseley}, {Breton},
  {Nikiel-Wroczy{\'n}ski}, {Sridhar}, {Stewart}, {Rowlinson}, {van der Horst},
  {Harwood}, {G{\"u}rkan}, {Carbone}, {Pandey-Pommier}, {Tasse}, {Scaife},
  {Pratley}, {Ferrari}, {Croston}, {Pandey}, {Jurusik}, \&
  {Mulcahy}}]{Kokotanekov2017}
{Kokotanekov}, G., {Wise}, M., {Heald}, G.~H., {et~al.} 2017, \aap, 605, A48

\bibitem[{{Kolokythas} {et~al.}(2020){Kolokythas}, {O'Sullivan}, {Giacintucci},
  {Worrall}, {Birkinshaw}, {Raychaudhury}, {Horellou}, {Intema}, \&
  {Loubser}}]{Kolokythas2020}
{Kolokythas}, K., {O'Sullivan}, E., {Giacintucci}, S., {et~al.} 2020, \mnras,
  496, 1471

\bibitem[{{Komissarov} \& {Gubanov}(1994)}]{Komissarov1994}
{Komissarov}, S.~S. \& {Gubanov}, A.~G. 1994, \aap, 285, 27

\bibitem[{{Konar} {et~al.}(2013){Konar}, {Hardcastle}, {Jamrozy}, \&
  {Croston}}]{Konar2013}
{Konar}, C., {Hardcastle}, M.~J., {Jamrozy}, M., \& {Croston}, J.~H. 2013,
  \mnras, 430, 2137

\bibitem[{{Laing}(1994)}]{Laing1994}
{Laing}, R.~A. 1994, in Astronomical Society of the Pacific Conference Series,
  Vol.~54, The Physics of Active Galaxies, ed. G.~V. {Bicknell}, M.~A.
  {Dopita}, \& P.~J. {Quinn}, 227

\bibitem[{{Maccagni} {et~al.}(2020){Maccagni}, {Murgia}, {Serra}, {Govoni},
  {Morokuma-Matsui}, {Kleiner}, {Buchner}, {J{\'o}zsa}, {Kamphuis},
  {Makhathini}, {Moln{\'a}r}, {Prokhorov}, {Ramaila}, {Ramatsoku}, {Thorat}, \&
  {Smirnov}}]{Maccagni2020}
{Maccagni}, F.~M., {Murgia}, M., {Serra}, P., {et~al.} 2020, \aap, 634, A9

\bibitem[{{McMullin} {et~al.}(2007){McMullin}, {Waters}, {Schiebel}, {Young},
  \& {Golap}}]{McMullin2007}
{McMullin}, J.~P., {Waters}, B., {Schiebel}, D., {Young}, W., \& {Golap}, K.
  2007, in Astronomical Society of the Pacific Conference Series, Vol. 376,
  Astronomical Data Analysis Software and Systems XVI, ed. R.~A. {Shaw},
  F.~{Hill}, \& D.~J. {Bell}, 127

\bibitem[{{McNamara} {et~al.}(2009){McNamara}, {Kazemzadeh}, {Rafferty},
  {B{\^\i}rzan}, {Nulsen}, {Kirkpatrick}, \& {Wise}}]{McNamara2009}
{McNamara}, B.~R., {Kazemzadeh}, F., {Rafferty}, D.~A., {et~al.} 2009, \apj,
  698, 594

\bibitem[{{McNamara} \& {Nulsen}(2007)}]{McNamaraNulsen2007}
{McNamara}, B.~R. \& {Nulsen}, P.~E.~J. 2007, \araa, 45, 117

\bibitem[{{McNamara} \& {Nulsen}(2012)}]{McNamaraNulsens2012}
{McNamara}, B.~R. \& {Nulsen}, P.~E.~J. 2012, New Journal of Physics, 14,
  055023

\bibitem[{{McNamara} {et~al.}(2005){McNamara}, {Nulsen}, {Wise}, {Rafferty},
  {Carilli}, {Sarazin}, \& {Blanton}}]{McNamara2005}
{McNamara}, B.~R., {Nulsen}, P.~E.~J., {Wise}, M.~W., {et~al.} 2005, \nat, 433,
  45

\bibitem[{{McNamara} {et~al.}(2000){McNamara}, {Wise}, {Nulsen}, {David},
  {Sarazin}, {Bautz}, {Markevitch}, {Vikhlinin}, {Forman}, {Jones}, \&
  {Harris}}]{McNamara2000}
{McNamara}, B.~R., {Wise}, M., {Nulsen}, P.~E.~J., {et~al.} 2000, \apjl, 534,
  L135

\bibitem[{{Meliani} {et~al.}(2008){Meliani}, {Keppens}, \&
  {Giacomazzo}}]{Meliani2008}
{Meliani}, Z., {Keppens}, R., \& {Giacomazzo}, B. 2008, \aap, 491, 321

\bibitem[{{Migliori} {et~al.}(2007){Migliori}, {Grandi}, {Palumbo}, {Brunetti},
  \& {Stanghellini}}]{Migliori2007}
{Migliori}, G., {Grandi}, P., {Palumbo}, G. G.~C., {Brunetti}, G., \&
  {Stanghellini}, C. 2007, \apj, 668, 203

\bibitem[{{Mitchell} {et~al.}(1976){Mitchell}, {Culhane}, {Davison}, \&
  {Ives}}]{Mitchell1976}
{Mitchell}, R.~J., {Culhane}, J.~L., {Davison}, P.~J.~N., \& {Ives}, J.~C.
  1976, \mnras, 175, 29P

\bibitem[{{Morganti}(2017)}]{Morganti2017}
{Morganti}, R. 2017, Nature Astronomy, 1, 596

\bibitem[{{Murgia} {et~al.}(1999){Murgia}, {Fanti}, {Fanti}, {Gregorini},
  {Klein}, {Mack}, \& {Vigotti}}]{Murgia1999}
{Murgia}, M., {Fanti}, C., {Fanti}, R., {et~al.} 1999, \aap, 345, 769

\bibitem[{{Myers} \& {Spangler}(1985)}]{MyersSpangler1985}
{Myers}, S.~T. \& {Spangler}, S.~R. 1985, \apj, 291, 52

\bibitem[{{Nulsen} {et~al.}(2005){Nulsen}, {McNamara}, {Wise}, \&
  {David}}]{Nulsen2005}
{Nulsen}, P.~E.~J., {McNamara}, B.~R., {Wise}, M.~W., \& {David}, L.~P. 2005,
  \apj, 628, 629

\bibitem[{{Offringa} {et~al.}(2014){Offringa}, {McKinley}, {Hurley-Walker},
  {Briggs}, {Wayth}, {Kaplan}, {Bell}, {Feng}, {Neben}, {Hughes}, {Rhee},
  {Murphy}, {Bhat}, {Bernardi}, {Bowman}, {Cappallo}, {Corey}, {Deshpand e},
  {Emrich}, {Ewall-Wice}, {Gaensler}, {Goeke}, {Greenhill}, {Hazelton},
  {Hindson}, {Johnston-Hollitt}, {Jacobs}, {Kasper}, {Kratzenberg}, {Lenc},
  {Lonsdale}, {Lynch}, {McWhirter}, {Mitchell}, {Morales}, {Morgan},
  {Kudryavtseva}, {Oberoi}, {Ord}, {Pindor}, {Procopio}, {Prabu}, {Riding},
  {Roshi}, {Shankar}, {Srivani}, {Subrahmanyan}, {Tingay}, {Waterson},
  {Webster}, {Whitney}, {Williams}, \& {Williams}}]{Offringa2014}
{Offringa}, A.~R., {McKinley}, B., {Hurley-Walker}, N., {et~al.} 2014, \mnras,
  444, 606

\bibitem[{{Orr} \& {Browne}(1982)}]{Orr1982}
{Orr}, M.~J.~L. \& {Browne}, I.~W.~A. 1982, \mnras, 200, 1067

\bibitem[{{Orr{\`u}} {et~al.}(2015){Orr{\`u}}, {van Velzen}, {Pizzo},
  {Yatawatta}, {Paladino}, {Iacobelli}, {Murgia}, {Falcke}, {Morganti}, {de
  Bruyn}, {Ferrari}, {Anderson}, {Bonafede}, {Mulcahy}, {Asgekar}, {Avruch},
  {Beck}, {Bell}, {van Bemmel}, {Bentum}, {Bernardi}, {Best}, {Breitling},
  {Broderick}, {Br{\"u}ggen}, {Butcher}, {Ciardi}, {Conway}, {Corstanje}, {de
  Geus}, {Deller}, {Duscha}, {Eisl{\"o}ffel}, {Engels}, {Frieswijk}, {Garrett},
  {Grie{\ss}meier}, {Gunst}, {Hamaker}, {Heald}, {Hoeft}, {van der Horst},
  {Intema}, {Juette}, {Kohler}, {Kondratiev}, {Kuniyoshi}, {Kuper}, {Loose},
  {Maat}, {Mann}, {Markoff}, {McFadden}, {McKay-Bukowski}, {Miley}, {Moldon},
  {Molenaar}, {Munk}, {Nelles}, {Paas}, {Pandey-Pommier}, {Pandey}, {Pietka},
  {Polatidis}, {Reich}, {R{\"o}ttgering}, {Rowlinson}, {Scaife},
  {Schoenmakers}, {Schwarz}, {Serylak}, {Shulevski}, {Smirnov}, {Steinmetz},
  {Stewart}, {Swinbank}, {Tagger}, {Tasse}, {Thoudam}, {Toribio}, {Vermeulen},
  {Vocks}, {van Weeren}, {Wijers}, {Wise}, \& {Wucknitz}}]{Orru2015}
{Orr{\`u}}, E., {van Velzen}, S., {Pizzo}, R.~F., {et~al.} 2015, \aap, 584,
  A112

\bibitem[{{Pacholczyk}(1970)}]{Pacholczyk1970}
{Pacholczyk}, A.~G. 1970, {Radio astrophysics. Nonthermal processes in galactic
  and extragalactic sources}

\bibitem[{{Perley} \& {Butler}(2013)}]{PerleyButler2013}
{Perley}, R.~A. \& {Butler}, B.~J. 2013, \apjs, 204, 19

\bibitem[{{Peterson} \& {Fabian}(2006)}]{PetersonFabian2006}
{Peterson}, J.~R. \& {Fabian}, A.~C. 2006, \physrep, 427, 1

\bibitem[{{Peterson} {et~al.}(2003){Peterson}, {Kahn}, {Paerels}, {Kaastra},
  {Tamura}, {Bleeker}, {Ferrigno}, \& {Jernigan}}]{Peterson2003}
{Peterson}, J.~R., {Kahn}, S.~M., {Paerels}, F.~B.~S., {et~al.} 2003, \apj,
  590, 207

\bibitem[{{Randall} {et~al.}(2011){Randall}, {Forman}, {Giacintucci}, {Nulsen},
  {Sun}, {Jones}, {Churazov}, {David}, {Kraft}, {Donahue}, {Blanton},
  {Simionescu}, \& {Werner}}]{Randall2011}
{Randall}, S.~W., {Forman}, W.~R., {Giacintucci}, S., {et~al.} 2011, \apj, 726,
  86

\bibitem[{{Randall} {et~al.}(2015){Randall}, {Nulsen}, {Jones}, {Forman},
  {Bulbul}, {Clarke}, {Kraft}, {Blanton}, {David}, {Werner}, {Sun}, {Donahue},
  {Giacintucci}, \& {Simionescu}}]{Randall2015}
{Randall}, S.~W., {Nulsen}, P.~E.~J., {Jones}, C., {et~al.} 2015, \apj, 805,
  112

\bibitem[{{Reynolds} {et~al.}(1996){Reynolds}, {Fabian}, {Celotti}, \&
  {Rees}}]{Reynolds1996}
{Reynolds}, C.~S., {Fabian}, A.~C., {Celotti}, A., \& {Rees}, M.~J. 1996,
  \mnras, 283, 873

\bibitem[{{Robitaille} \& {Bressert}(2012)}]{Robitaille2012}
{Robitaille}, T. \& {Bressert}, E. 2012, {APLpy: Astronomical Plotting Library
  in Python}

\bibitem[{{Scaife} \& {Heald}(2012)}]{ScaifeHeald2012}
{Scaife}, A. M.~M. \& {Heald}, G.~H. 2012, \mnras, 423, L30

\bibitem[{{Scheuer} \& {Williams}(1968)}]{Scheuer1968}
{Scheuer}, P.~A.~G. \& {Williams}, P.~J.~S. 1968, \araa, 6, 321

\bibitem[{{Serlemitsos} {et~al.}(1977){Serlemitsos}, {Smith}, {Boldt}, {Holt},
  \& {Swank}}]{Serlemitsos1977}
{Serlemitsos}, P.~J., {Smith}, B.~W., {Boldt}, E.~A., {Holt}, S.~S., \&
  {Swank}, J.~H. 1977, \apjl, 211, L63

\bibitem[{{Shimwell} {et~al.}(2017){Shimwell}, {R{\"o}ttgering}, {Best},
  {Williams}, {Dijkema}, {de Gasperin}, {Hardcastle}, {Heald}, {Hoang},
  {Horneffer}, {Intema}, {Mahony}, {Mandal}, {Mechev}, {Morabito}, {Oonk},
  {Rafferty}, {Retana-Montenegro}, {Sabater}, {Tasse}, {van Weeren},
  {Br{\"u}ggen}, {Brunetti}, {Chy{\.z}y}, {Conway}, {Haverkorn}, {Jackson},
  {Jarvis}, {McKean}, {Miley}, {Morganti}, {White}, {Wise}, {van Bemmel},
  {Beck}, {Brienza}, {Bonafede}, {Calistro Rivera}, {Cassano}, {Clarke},
  {Cseh}, {Deller}, {Drabent}, {van Driel}, {Engels}, {Falcke}, {Ferrari},
  {Fr{\"o}hlich}, {Garrett}, {Harwood}, {Heesen}, {Hoeft}, {Horellou},
  {Israel}, {Kapi{\'n}ska}, {Kunert-Bajraszewska}, {McKay}, {Mohan},
  {Orr{\'u}}, {Pizzo}, {Prandoni}, {Schwarz}, {Shulevski}, {Sipior}, {Smith},
  {Sridhar}, {Steinmetz}, {Stroe}, {Varenius}, {van der Werf}, {Zensus}, \&
  {Zwart}}]{Shimwell2017}
{Shimwell}, T.~W., {R{\"o}ttgering}, H.~J.~A., {Best}, P.~N., {et~al.} 2017,
  \aap, 598, A104

\bibitem[{{Shimwell} {et~al.}(2019){Shimwell}, {Tasse}, {Hardcastle}, {Mechev},
  {Williams}, {Best}, {R{\"o}ttgering}, {Callingham}, {Dijkema}, {de Gasperin},
  {Hoang}, {Hugo}, {Mirmont}, {Oonk}, {Prandoni}, {Rafferty}, {Sabater},
  {Smirnov}, {van Weeren}, {White}, {Atemkeng}, {Bester}, {Bonnassieux},
  {Br{\"u}ggen}, {Brunetti}, {Chy{\.z}y}, {Cochrane}, {Conway}, {Croston},
  {Danezi}, {Duncan}, {Haverkorn}, {Heald}, {Iacobelli}, {Intema}, {Jackson},
  {Jamrozy}, {Jarvis}, {Lakhoo}, {Mevius}, {Miley}, {Morabito}, {Morganti},
  {Nisbet}, {Orr{\'u}}, {Perkins}, {Pizzo}, {Schrijvers}, {Smith}, {Vermeulen},
  {Wise}, {Alegre}, {Bacon}, {van Bemmel}, {Beswick}, {Bonafede}, {Botteon},
  {Bourke}, {Brienza}, {Calistro Rivera}, {Cassano}, {Clarke}, {Conselice},
  {Dettmar}, {Drabent}, {Dumba}, {Emig}, {En{\ss}lin}, {Ferrari}, {Garrett},
  {G{\'e}nova-Santos}, {Goyal}, {G{\"u}rkan}, {Hale}, {Harwood}, {Heesen},
  {Hoeft}, {Horellou}, {Jackson}, {Kokotanekov}, {Kondapally},
  {Kunert-Bajraszewska}, {Mahatma}, {Mahony}, {Mandal}, {McKean}, {Merloni},
  {Mingo}, {Miskolczi}, {Mooney}, {Nikiel-Wroczy{\'n}ski}, {O'Sullivan},
  {Quinn}, {Reich}, {Roskowi{\'n}ski}, {Rowlinson}, {Savini}, {Saxena},
  {Schwarz}, {Shulevski}, {Sridhar}, {Stacey}, {Urquhart}, {van der Wiel},
  {Varenius}, {Webster}, \& {Wilber}}]{Shimwell2019}
{Shimwell}, T.~W., {Tasse}, C., {Hardcastle}, M.~J., {et~al.} 2019, \aap, 622,
  A1

\bibitem[{{Shulevski} {et~al.}(2017){Shulevski}, {Morganti}, {Harwood},
  {Barthel}, {Jamrozy}, {Brienza}, {Brunetti}, {R{\"o}ttgering}, {Murgia},
  {White}, {Croston}, \& {Br{\"u}ggen}}]{Shulevski2017}
{Shulevski}, A., {Morganti}, R., {Harwood}, J.~J., {et~al.} 2017, \aap, 600,
  A65

\bibitem[{{Smirnov} \& {Tasse}(2015)}]{SmirnovTass2015}
{Smirnov}, O.~M. \& {Tasse}, C. 2015, \mnras, 449, 2668

\bibitem[{{Swarup}(1990)}]{Swarup1990}
{Swarup}, G. 1990, Indian Journal of Radio and Space Physics, 19, 493

\bibitem[{{Tadhunter}(2016)}]{Tadhunter2016}
{Tadhunter}, C. 2016, \aapr, 24, 10

\bibitem[{{Tasse}(2014{\natexlab{a}})}]{Tasse2014a}
{Tasse}, C. 2014{\natexlab{a}}, arXiv e-prints, arXiv:1410.8706

\bibitem[{{Tasse}(2014{\natexlab{b}})}]{Tasse2014b}
{Tasse}, C. 2014{\natexlab{b}}, \aap, 566, A127

\bibitem[{{Tasse} {et~al.}(2018){Tasse}, {Hugo}, {Mirmont}, {Smirnov},
  {Atemkeng}, {Bester}, {Hardcastle}, {Lakhoo}, {Perkins}, \&
  {Shimwell}}]{Tasse2018}
{Tasse}, C., {Hugo}, B., {Mirmont}, M., {et~al.} 2018, \aap, 611, A87

\bibitem[{{Tribble}(1993)}]{Tribble1993}
{Tribble}, P.~C. 1993, \mnras, 261, 57

\bibitem[{{Turner} \& {Shabala}(2015)}]{TurnerShabala2015}
{Turner}, R.~J. \& {Shabala}, S.~S. 2015, \apj, 806, 59

\bibitem[{{Turner} {et~al.}(2018){Turner}, {Shabala}, \& {Krause}}]{Turner2018}
{Turner}, R.~J., {Shabala}, S.~S., \& {Krause}, M. G.~H. 2018, \mnras, 474,
  3361

\bibitem[{{van der Walt} {et~al.}(2011){van der Walt}, {Colbert}, \&
  {Varoquaux}}]{vanderwalt2011}
{van der Walt}, S., {Colbert}, S.~C., \& {Varoquaux}, G. 2011, Computing in
  Science and Engineering, 13, 22

\bibitem[{{van Diepen} {et~al.}(2018){van Diepen}, {Dijkema}, \&
  {Offringa}}]{DPPP2018}
{van Diepen}, G., {Dijkema}, T.~J., \& {Offringa}, A. 2018, {DPPP: Default
  Pre-Processing Pipeline}

\bibitem[{{van Haarlem} {et~al.}(2013){van Haarlem}, {Wise}, {Gunst}, {Heald},
  {McKean}, {Hessels}, {de Bruyn}, {Nijboer}, {Swinbank}, {Fallows},
  {Brentjens}, {Nelles}, {Beck}, {Falcke}, {Fender}, {H{\"o}randel},
  {Koopmans}, {Mann}, {Miley}, {R{\"o}ttgering}, {Stappers}, {Wijers},
  {Zaroubi}, {van den Akker}, {Alexov}, {Anderson}, {Anderson}, {van Ardenne},
  {Arts}, {Asgekar}, {Avruch}, {Batejat}, {B{\"a}hren}, {Bell}, {Bell}, {van
  Bemmel}, {Bennema}, {Bentum}, {Bernardi}, {Best}, {B{\^\i}rzan}, {Bonafede},
  {Boonstra}, {Braun}, {Bregman}, {Breitling}, {van de Brink}, {Broderick},
  {Broekema}, {Brouw}, {Br{\"u}ggen}, {Butcher}, {van Cappellen}, {Ciardi},
  {Coenen}, {Conway}, {Coolen}, {Corstanje}, {Damstra}, {Davies}, {Deller},
  {Dettmar}, {van Diepen}, {Dijkstra}, {Donker}, {Doorduin}, {Dromer}, {Drost},
  {van Duin}, {Eisl{\"o}ffel}, {van Enst}, {Ferrari}, {Frieswijk}, {Gankema},
  {Garrett}, {de Gasperin}, {Gerbers}, {de Geus}, {Grie{\ss}meier}, {Grit},
  {Gruppen}, {Hamaker}, {Hassall}, {Hoeft}, {Holties}, {Horneffer}, {van der
  Horst}, {van Houwelingen}, {Huijgen}, {Iacobelli}, {Intema}, {Jackson},
  {Jelic}, {de Jong}, {Juette}, {Kant}, {Karastergiou}, {Koers}, {Kollen},
  {Kondratiev}, {Kooistra}, {Koopman}, {Koster}, {Kuniyoshi}, {Kramer},
  {Kuper}, {Lambropoulos}, {Law}, {van Leeuwen}, {Lemaitre}, {Loose}, {Maat},
  {Macario}, {Markoff}, {Masters}, {McFadden}, {McKay-Bukowski}, {Meijering},
  {Meulman}, {Mevius}, {Middelberg}, {Millenaar}, {Miller-Jones}, {Mohan},
  {Mol}, {Morawietz}, {Morganti}, {Mulcahy}, {Mulder}, {Munk}, {Nieuwenhuis},
  {van Nieuwpoort}, {Noordam}, {Norden}, {Noutsos}, {Offringa}, {Olofsson},
  {Omar}, {Orr{\'u}}, {Overeem}, {Paas}, {Pand ey-Pommier}, {Pandey}, {Pizzo},
  {Polatidis}, {Rafferty}, {Rawlings}, {Reich}, {de Reijer}, {Reitsma},
  {Renting}, {Riemers}, {Rol}, {Romein}, {Roosjen}, {Ruiter}, {Scaife}, {van
  der Schaaf}, {Scheers}, {Schellart}, {Schoenmakers}, {Schoonderbeek},
  {Serylak}, {Shulevski}, {Sluman}, {Smirnov}, {Sobey}, {Spreeuw}, {Steinmetz},
  {Sterks}, {Stiepel}, {Stuurwold}, {Tagger}, {Tang}, {Tasse}, {Thomas},
  {Thoudam}, {Toribio}, {van der Tol}, {Usov}, {van Veelen}, {van der Veen},
  {ter Veen}, {Verbiest}, {Vermeulen}, {Vermaas}, {Vocks}, {Vogt}, {de Vos},
  {van der Wal}, {van Weeren}, {Weggemans}, {Weltevrede}, {White}, {Wijnholds},
  {Wilhelmsson}, {Wucknitz}, {Yatawatta}, {Zarka}, {Zensus}, \& {van
  Zwieten}}]{vanHaarlem2013}
{van Haarlem}, M.~P., {Wise}, M.~W., {Gunst}, A.~W., {et~al.} 2013, \aap, 556,
  A2

\bibitem[{{van Weeren} {et~al.}(2019){van Weeren}, {de Gasperin}, {Akamatsu},
  {Br{\"u}ggen}, {Feretti}, {Kang}, {Stroe}, \& {Zandanel}}]{vanWeeren2019}
{van Weeren}, R.~J., {de Gasperin}, F., {Akamatsu}, H., {et~al.} 2019, \ssr,
  215, 16

\bibitem[{{Vantyghem} {et~al.}(2014){Vantyghem}, {McNamara}, {Russell}, {Main},
  {Nulsen}, {Wise}, {Hoekstra}, \& {Gitti}}]{Vantyghem2014}
{Vantyghem}, A.~N., {McNamara}, B.~R., {Russell}, H.~R., {et~al.} 2014, \mnras,
  442, 3192

\bibitem[{{Virtanen} {et~al.}(2020){Virtanen}, {Gommers}, {Oliphant},
  {Haberland}, {Reddy}, {Cournapeau}, {Burovski}, {Peterson}, {Weckesser},
  {Bright}, {van der Walt}, {Brett}, {Wilson}, {Millman}, {Mayorov}, {Nelson},
  {Jones}, {Kern}, {Larson}, {Carey}, {Polat}, {Feng}, {Moore}, {VanderPlas},
  {Laxalde}, {Perktold}, {Cimrman}, {Henriksen}, {Quintero}, {Harris},
  {Archibald}, {Ribeiro}, {Pedregosa}, {van Mulbregt}, \& {SciPy 1. 0
  Contributors}}]{Scipy2020}
{Virtanen}, P., {Gommers}, R., {Oliphant}, T.~E., {et~al.} 2020, Nature
  Methods, 17, 261

\bibitem[{{Wang} {et~al.}(2011){Wang}, {Knigge}, {Croston}, \&
  {Pavlovski}}]{Wang2011}
{Wang}, Y., {Knigge}, C., {Croston}, J.~H., \& {Pavlovski}, G. 2011, \mnras,
  418, 1138

\bibitem[{{Wise} {et~al.}(2007){Wise}, {McNamara}, {Nulsen}, {Houck}, \&
  {David}}]{Wise2007}
{Wise}, M.~W., {McNamara}, B.~R., {Nulsen}, P.~E.~J., {Houck}, J.~C., \&
  {David}, L.~P. 2007, \apj, 659, 1153

\end{thebibliography}

\onecolumn
\appendix
\section{$\alpha_{\rm inj} = 1.5$}\label{sec:appendix}
We report the fitting results obtained setting the injection index to $\alpha_{\rm inj}=1.5$, which gives the lowest chi-squared value to the models fit. The results of the fit computed for both the Tribble and CIoff models, together with the ages corrected for the presence of adiabatic losses, are summarised in Table \ref{tab:conf}.

\begin{table*}[h]
    \centering
    \renewcommand\arraystretch{1.2}
    \caption{Tribble and CIoff fitting results and correction for adiabatic expansion.}
    \begin{tabular}{cc|ccccc|ccccc}\hline
      \multicolumn{12}{c}{$\alpha_{\rm inj}= 1.5$ \qquad $B_{\rm eq} = 15\ \rm{\mu G}$}\\\hline 
    \multicolumn{2}{c}{} &\multicolumn{5}{c}{Tribble model} &\multicolumn{5}{c}{CIoff model}\\
      Region &Age &$\nu_{\rm break}$ &$t$ &$\nu^{\rm corr}_{\rm break}$ &$t^{\rm corr}$ &$\chi^2_{\rm red}$ &$\nu_{\rm break}$ &$t$ &$\nu^{\rm corr}_{\rm break}$ &$t^{\rm corr}$ &$\chi^2_{\rm red}$\\
&&[MHz] &[Myr] &[MHz] &[Myr] &[MHz] &[Myr] &[MHz] &[Myr] \\\hline
Northern outer lobe &Tot &358 &$42\pm3$ &931 &$26\pm2$ &0.48 &188 &$47\pm7$ &11 &$238\pm28$ &1.11\\
&Off &1170 &$19\pm2$ &3042 &$14\pm1$ &3.43 &1160 &$19\pm1$ &70 &$94\pm4$ &$-$\\
&On &$-$&$23\pm5$&$-$&$12\pm3$ &$-$ &$-$ &$28\pm8$ &$-$&$144\pm32$ &$-$\\
Southern outer lobe &Tot &424 &$38\pm2$ &1230 &$22\pm1$ &0.35 &153 &$52\pm13$ &9 &$263\pm52$ &0.73\\
&Off &741 &$29\pm2$ &2149 &$17\pm1$ &0.97 &1240 &$18\pm1$ &74 &$92\pm4$ &$-$\\
&On &$-$&$9\pm4$&$-$&$5\pm2$ &$-$ &$-$ &$34\pm14$ &$-$&$171\pm56$ &$-$\\
Intermediate lobe &Tot &743 &$30\pm2$ &1040 &$24\pm2$ &0.72 &$-$&$-$ &$-$&$-$ &$-$\\\hline
    \end{tabular}
    \tablefoot{The break frequency ($\nu_{\rm break}$) and the synchrotron age ($t$) derived from the fitting model, the adiabatic-loss corrected break frequency ($\nu_{\rm break}^{\rm corr}$) and synchrotron age ($t^{\rm corr}$), and finally, the reduced chi-squared of the fit ($\chi^2_{\rm red}$). These parameters are derived for both the total age (Tot), the dying phase (Off), and the active phase (On) of the lobes using two different models (Tribble and CIoff), setting $\alpha_{\rm inj}=1.5$ and $B_{\rm eq}=15\ \mu G$).}
    \label{tab:conf}
\end{table*}

\end{document}